\documentclass{aa}  

\usepackage{graphicx}
\usepackage{txfonts}
\usepackage[colorlinks=true,linkcolor=blue,citecolor=blue,urlcolor=blue]{hyperref}
\usepackage[separate-uncertainty=true]{siunitx}

\begin{document} 

   \title{Connecting stellar and galactic scales: Energetic feedback from stellar wind bubbles to supernova remnants}

   \titlerunning{Connecting stellar and galactic scales}

        \author{Yvonne A. Fichtner\inst{1}
                \and
                Jonathan Mackey\inst{2}
                \and
                Luca Grassitelli\inst{1}
                \and
                Emilio Romano-D\'{\i}az\inst{1}
                \and
                Cristiano Porciani\inst{1,3,4,5}
        }
        
        \institute{Argelander-Institut für Astronomie, Auf dem Hügel 71, D-53121 Bonn, Germany\\
                \email{yfichtner@astro.uni-bonn.de}
                \and
                Dublin Institute for Advanced Studies, Astronomy and Astrophysics Section, DIAS Dunsink Observatory, Dublin D15 XR2R, Ireland
        \and
        SISSA, International School for Advanced Studies, Via Bonomea 265, 34136 Trieste, TS, Italy
        \and
        Dipartimento di Fisica – Sezione di Astronomia, Universit\`a di Trieste, Via Tiepolo 11, 34131 Trieste, Italy
        \and
        IFPU, Institute for Fundamental Physics of the Universe, via Beirut 2, 34151 Trieste, Italy
        }

   \date{Received XXXX ; accepted YYYY}

    \authorrunning{Yvonne A. Fichtner et al.}
    
 
        \abstract
        {Energy and momentum feedback from stars is a key element in models of galaxy formation and interstellar medium (ISM) dynamics, but resolving the relevant length scales in order to directly include this feedback remains beyond the reach of current-generation simulations.}
        {We aim to constrain the energy feedback of winds, photoionisation, and supernovae (SNe) from massive stars.}
        {We measure the thermal and kinetic energy imparted to the ISM on various length scales, which we calculate from high-resolution 1D radiation-hydrodynamics simulations. Our grid of simulations covers a broad range of densities, metallicities, and state-of-the-art evolutionary models of single and binary stars.}
        {A single star or binary system can carve a cavity of tens of parsecs (pc) in size into the surrounding medium. During the pre-SN phase, post-main sequence stellar winds and photoionisation dominate. While SN explosions dominate the total energy budget, the pre-SN feedback is of great importance by reducing the circumstellar gas density and delaying the onset of radiative losses in the SN remnant. Contrary to expectations, the metallicity dependence of the stellar wind has little effect on the cumulative energy imparted by feedback to the ISM; the only requirement is the existence of a sufficient level of pre-SN radiative and mechanical feedback. The ambient medium density determines how much and when feedback energy reaches distances of $\gtrsim 10-20$\,pc and affects the division between kinetic and thermal feedback.}
        {Our results can be used as a subgrid model for feedback in large-scale simulations of galaxies. The results reinforce that the uncertain mapping of stellar evolution sequences to SN explosion energy is very important for determining the overall feedback energy from a stellar population.}
        

   \maketitle
%



\section{Introduction}
There is a consensus that the feedback from stars is key to explaining the physical properties of the observed galaxy population \citep[e.g.][]{Naab2017}.
In fact, as seen in numerical simulations, the injection of energy and metals from stars into the surrounding medium regulates the efficiency of star formation \citep[SF; e.g.][]{Hopkins2011, Hopkins2014, Kimm2017, Marinacci2019} and prevents the assembly of overly dense stellar bulges \citep[e.g.][]{AgertzKravtsov2015}.
Feedback also enriches the circumgalactic medium with products of stellar nucleosynthesis \citep[e.g.][]{Muratov2017, Nelson2019}.

Early models of stellar feedback only accounted for the thermal energy from core-collapse supernovae \citep[CCSNe; e.g.][]{Dekel1986,Katz1992}. 
However, other feedback processes that occur before the SN explosions (e.g. stellar winds and photoionisation) should also make important contributions \citep[e.g.][]{ Dib2011,Hopkins2011,Brook2012,Stinson2012}.  
Evidence for this pre-SN feedback comes from observations that gas is dispersed from the birth places of stars before the first SN explosions take place \citep[e.g.][]{Mathews1966,Castor1975b,Barnes2020, Chevance2020}.
Theoretical models suggest that preprocessing the immediate stellar environment could also strengthen SN feedback \citep[e.g.][]{Agertz2013, Geen2015, Fierlinger2016, Hopkins2018, Kannan2020, Lucas2020}. 

What makes modelling stellar feedback very challenging is the wide range of spatial and temporal scales that regulate the physical processes \citep[see e.g.][]{Geen2023_Workshop}{}{}. 
For this reason, different approaches are used to investigate feedback effects on different scales. On subgalactic scales, 
pre-SN feedback alone can already have a strong effect on the surrounding medium \citep[e.g.][]{Rogers2013,Fierlinger2016,Gatto2017}, with stellar winds creating cavities of tens of parsecs (pc) in size inside H\,\textsc{ii} regions \citep{Dale2014}.
The strong metallicity dependence of the stellar wind strength \citep{vink2021r} implies that early feedback may be significantly less effective at very low metallicity, but this has not yet been thoroughly explored theoretically or confirmed observationally.

It has been observed that a single stellar system can have a strong influence on its surroundings. For instance, \citet{Ramachandran2019} showed that in the `wing' of the Small Magellanic Cloud (SMC), a single Wolf-Rayet (WR) binary system is dominating the pre-SN feedback over the combined feedback of all OB stars. This highlights the importance of post-main sequence phases when determining stellar feedback.

In addition to improvements in the considered feedback processes, the subgrid injection model has also been improved since its initial implementation, resulting in the existence of different schemes.
In some models, the energy is injected completely in the thermal component \citep[e.g.][]{Katz1992, Stinson2006, DallaVecchia2012, Schaye2015}{}{},
while other models inject kinetic energy and momentum \citep[e.g.][]{DallaVecchia2008,Dubois2008,Gatto2015,Kimm2015}{}{},
a mixture of both components \citep[e.g.][]{Dave2017,Hopkins2018,Fichtner2022,Chaikin2023},{}{}
or switching between different methods \citep[e.g.][]{Agertz2020}.
The choice of the subgrid feedback scheme is often driven by numerical considerations.
The strength of stellar feedback is usually either derived from libraries, for example \textsc{Starburst99} \citep{Leitherer1999,Vazquez2005,Leitherer2010,Leitherer2014}  or \textsc{BPASS} \citep[Binary Population and Spectral Synthesis,][]{Eldridge2017,Stanway2018}, or calibrated by comparing 
the properties of simulated and observed galaxies.

The direct use of results from population synthesis codes neglects the fact that the feedback is altered by the unresolved surrounding medium with respect to the stellar ejection.
Some of the feedback energy is likely dissipated \citep[e.g.][]{GarciaSeguraLanger1996, GarciaSeguraMacLow1996} at subresolution scales. 
Cosmological full-box simulations can achieve spatial resolution of the order of $100$ pc or higher \citep[e.g.][]{Feldmann2023}, while simulations that zoom onto a selected region can achieve higher resolutions of a few pc or even subpc resolutions \citep[e.g.][]{KimmCen2016,KimWise2019,Agertz2020,Gutcke2022,Calura2022}{}{}. However, energy dissipation at
pc and subpc scale cannot be modelled correctly in these simulations when they are not resolved. 
In addition,
energy dissipation can be due to the variable stellar wind during the lifetime of a star \citep{Vink2001, vink2021r}, for instance when the fast WR wind catches up to the slow and dense wind of the red supergiant or luminous blue variable phase.
However, the strength of the energy dissipation is unclear. \citet{Rey-Raposo2017} and \citet{Lancaster2021} argue that most of the feedback energy is radiated away, while \citet{Rosen2021} find that radiative losses are insignificant. 

The interstellar medium (ISM) has a complex, multiphase structure \citep[e.g.][]{Cox1974, McKee1977, MacLow1988, Rosen1993, Silich2008}, which leads to deviations from spherical symmetry in the evolution of stellar-feedback bubbles, as each bubble can expand further into regions of lower density \citep[e.g.][]{ZamoraAviles2019}. 
Finger-like structures \citep[e.g.][]{GarciaSeguraFranco1996, Dale2005} and champagne flows \citep[e.g.][]{TenorioTagle1979, Franco1990, Geen2023} can also be triggered by radiation-hydrodynamical instabilities. 
These processes can only be modelled in multidimensional simulations. However, 3D simulations with multiphase structures are numerically expensive, which limits these studies to one or a few models.

In this work, we investigate the energy deposition from the feedback of massive stars at various distances from the stellar sources through high-resolution 1D radiation-hydrodynamics simulations. These simulations allow us to better estimate the radiative losses occurring at small scales and to begin to bridge the gap between stellar and galactic length scales.
To this end, we generated a grid of over $2500$ simulations of stellar feedback from single and binary stars in a homogeneous ISM. We include the feedback from stellar winds, photoionisation, and CCSNe. Simulations were performed with the grid-based fluid dynamics code \textsc{Pion} \citep{Mackey2021}.
As input for the \textsc{Pion} simulations, we used the stellar models of massive stars described in \citet{Fichtner2022}.
As well as single stars, the grid includes models of binary systems, as most massive stars are born as binaries \citep[][]{Sana2012,Sana2013}{}{}. We employed models with different initial surface rotation velocity of single stars and binary systems with different orbital periods. We investigated five different subsolar metallicities, with the aim being to replicate the conditions found in high-redshift galaxies. 
We used these simulations to investigate how much energy reaches different radii at different ambient densities. This is the energy budget that is available to drive feedback in galaxy simulations that do not resolve pc scales. Additionally, we analysed the ratio of kinetic to total energy, which is often a free parameter in subgrid feedback models.

The paper is organised as follows: In Sect. \ref{sec:Stellar_models} we introduce the grid of stellar models. In Sect. \ref{sec:Numerical methods}, the numerical methods and initial conditions of the \textsc{Pion} simulations are described. We present our analysis in Sect. \ref{sec:Results}, and discuss the implications of our findings in Sect. \ref{sec:Discussion}. We then review our main conclusions in Sect. \ref{sec:Conclusions}.

\section{Stellar models}
\label{sec:Stellar_models}
\subsection{Stellar grid}
\label{sec:Stellar_models_life}
\begin{table*}
    \centering
    \caption{Naming convention of the different sequences of stellar models in the grid. }
    \label{tab:Stellar_naming}
    \begin{tabular}{ccc}
        \hline
        \noalign{\smallskip}
        System  & Naming convection & Example \\ 
        \noalign{\smallskip}  
        \hline  
        \noalign{\smallskip}
        Single  &  single star `s' and initial rotation velocity in $\SI{e2}{\kilo\meter\per\second}$ 
        & s$3$: single star rotating with \SI{300}{\kilo\meter\per\second} \\
        Binary  & binary system `b' and orbital period $P$ in $\log_{10}(P\mathrm{\,[days]})$,
        Note: $\mathrm{b}0.6=\mathrm{b}.6$
        & b$1$: binary system with $P = 10^{1}$ days\\ 
       \noalign{\smallskip}
       \hline
    \end{tabular}   
\end{table*}
We adopt the stellar models from \citet{Fichtner2022}, whose main properties we summarise here. The stellar models are computed with the stellar-evolution code Modules for Experiments in Stellar Astrophysics \citep[\textsc{Mesa}, v.8845,][]{Paxton2011, Paxton2013, Paxton2015, Paxton2018, Paxton2019}. 
The grid contains models of single, rotating massive stars as well as models of stars in binary systems. The inclusion of binary systems is of paramount importance as the majority of massive stars are born in binaries \citep{Sana2012, Sana2013}. Binary stars may experience phases of mass transfer when they fill their Roche lobes. The binary evolution creates evolutionary pathways and spectroscopic appearances which are not achievable for single stars \citep[e.g.][]{deMarco2017}{}{}.
We use stellar models in the mass range $\log_{10}(M_\textrm{ini}/\textrm{M}_\odot)=0.9-2.2$ in steps of $\log_{10}(M_\textrm{ini}/\textrm{M}_\odot)=0.1$ for single stars and in the range $\log_{10}(M_\textrm{ini}/\textrm{M}_\odot)=1.0-2.2$ in steps of $\log_{10}(M_\textrm{ini}/\textrm{M}_\odot)=0.2$ for the primaries of the binary systems. For the single stars, the grid spans a wide range of initial surface rotation velocities, namely $0$ to \SI{600}{\kilo\meter\per\second} in steps of \SI{100}{\kilo\meter\per\second}.  
For the binary systems, the initial rotation velocity is set to give tidal locking. 
Four orbital periods ($P = 10^{0.6}$, $10^1$, $10^2$, and $10^3$ days) are considered while, for simplicity, a fixed mass ratio of $0.6$ is assumed (see Table \ref{tab:Stellar_naming} for the naming convention used in this paper). 

In this work, we investigate the subsolar metallicities $Z=0.0001$, $0.0004$, $0.0008$, $0.001$ and $0.004$, which are representative of the metallicities observed in high-redshift and/or low-mass  
galaxies \citep[e.g.][]{Maiolino2008,Madau2014}.
Throughout this work, we state metallicities in absolute units and not relative to the solar metallicity.
These metallicities correspond to the lower metallicities used in the binary grid, adding the single stellar models for the metallicity \num{0.0008} with respect to \citet{Fichtner2022}.
As in the previous paper, we stop the evolution of the stellar models at core He-exhaustion. In the remainder of the section we highlight the difference with respect to \citet{Fichtner2022}.

In contrast to other stellar input properties, the terminal wind velocity $\upsilon_\infty$ is not provided directly by \textsc{Mesa}.
We calculate $\upsilon_\infty$ as in equation (2) of \citet{Fichtner2022}, which was based on 
\citet{Leitherer1992} and \citet{Puls2008}, but we modify the multiplicative prefactor between $\upsilon_\infty$ and the escape velocity.
It is changed to follow the prescription of \citet{Eldridge2006}, 
except that a factor of $2.0$ is used for WR stars (identified by a temperature above \SI{3e4}{K} and a hydrogen abundance below $0.6$). 

For binary systems the individual stellar wind and radiation output must be combined into averaged, spherically symmetry quantities that can be input into 1D simulations, achieved by weighting the energy of each wind according to solid angle (see below).
We add the mass- and metals-loss rates of the two stellar models and assume the temperature of the hotter star for the combined object. The luminosities are added with a correction factor so that the combined ionising photon luminosity is the sum of the two sources.
To calculate the terminal wind velocity, 
we add the kinetic energy of the wind of the two companions, where
we take into account energy dissipation due to wind-wind interactions in the binary system
with an improved method (described in the following) compared to \citet{Fichtner2022}.
We assume that, at large distances from the system, only a fraction of the solid angle allows for the freely expanding wind of each component of the binary system.
Based on the models by \citet{Eichler1993} and \citet[][but see also \citealt{Pittard2018}]{Girard1987}, the region filled by the weaker stellar wind can be approximated by a cone.
Half of the opening angle of that cone ($\theta$) can be estimated numerically from the stellar mass-loss rate $\dot{M}$ and terminal wind velocity $\upsilon_\infty$ via: 
\begin{equation}
    \theta=2.1\,\left(1-\frac{\nu^{2/5}}{4}\right)\nu^{1/3}\,,
\end{equation}
with 
\begin{equation}
    \nu=\frac{\dot{M_\textrm{2}}\upsilon_{\infty,\textrm{2}}}{\dot{M_\textrm{1}}\upsilon_{\infty,\textrm{1}}}\,,
\end{equation} 
the indices $1$ and $2$ indicate the component with the stronger (given by $\dot{M}\upsilon_\infty$) and weaker wind respectively, so that $\nu\leq1$. 
The normalised solid angle of the cone of the weaker model $2$ is then
\begin{equation}
    f_\textrm{ww,2}=0.5\,(1-\cos{\theta})\,, \\
\end{equation}
and the solid angle of the stronger wind model $1$ is, consequently:
\begin{equation}
    f_\textrm{ww,1}=1-f_\textrm{ww,2}.
\end{equation}
By doing so, we find that the stellar model with the wind of higher momentum covers a larger solid angle, mimicking the  increased momentum of its freely expanding outflow at large scales while accounting for energy dissipation in the region of colliding winds.

\subsection{Supernova}
\label{sec:Stellar_models_SN}
\begin{figure}
    \centering
    \includegraphics[width=1\columnwidth]{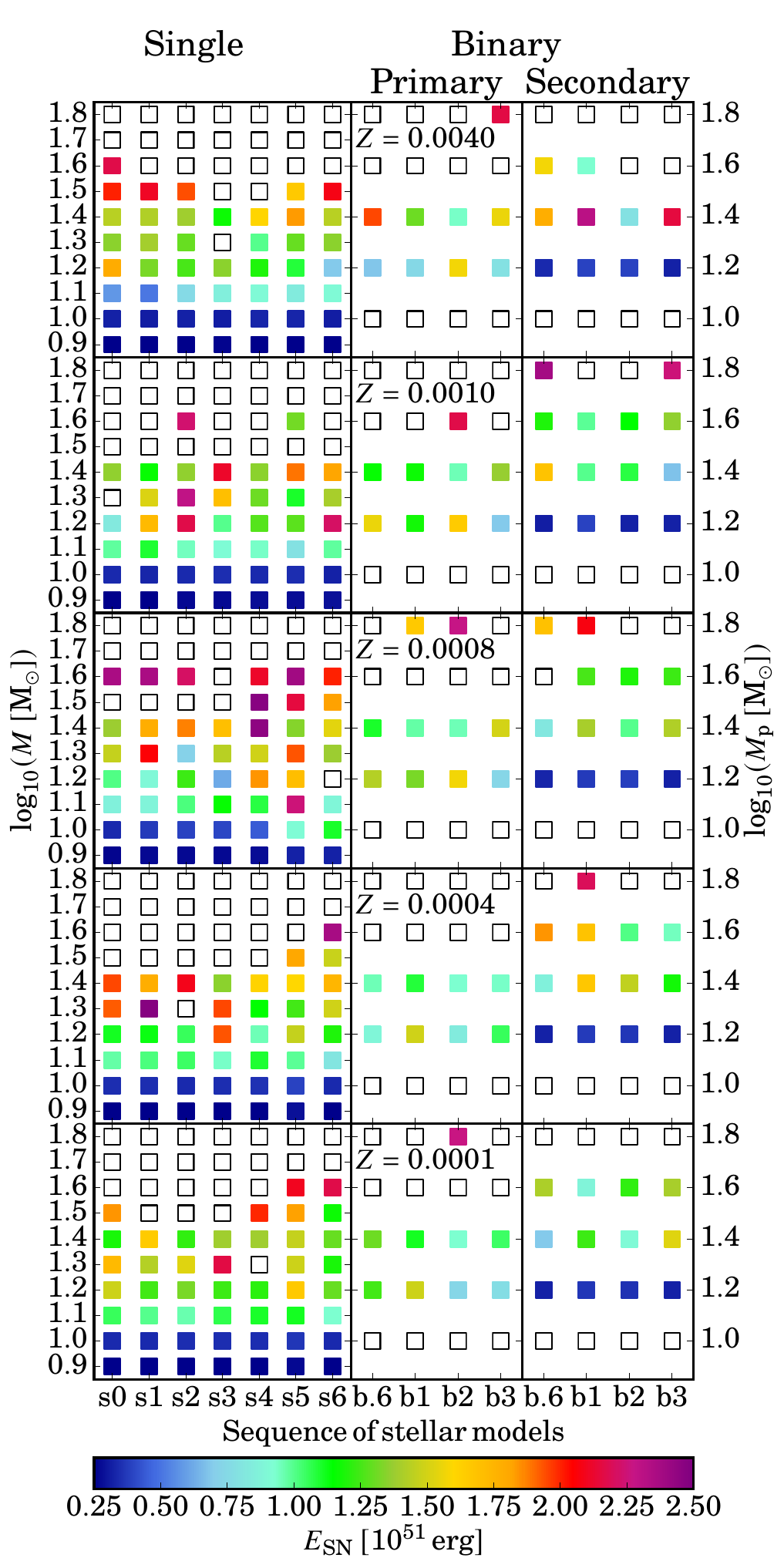}
    \caption{Explosion energy of each stellar model if it has a CCSN, where the colour of the squares indicates the SN energy. White squares indicate models that do not have a CCSN. The left column shows the single models, while the two right columns show the primary and secondary models of binary systems, respectively. Each row shows one metallicity. 
    The x-axis shows the different names according to Table \ref{tab:Stellar_naming}.
    The initial stellar masses are shown on the y-axis; we note that the binary grid has a coarser sampling in initial mass than the single grid. The highest initial masses are not shown as they have no CCSN.}
    \label{fig:Stellar_SN}
\end{figure}
In order to account for the variety of CCSN types, we employ the compactness parameter to infer which stellar models could explode as a CCSN. This parameter is a measure of the 
density (and binding energy) of the stellar core at bounce, evaluated at mass coordinate \SI{2.5}{M_{\sun}}, defined as $2.5 / R_{2.5}$, where $R_{2.5}$ is the radius at bounce that encloses \SI{2.5}{M_{\sun}} in units of $10^3$\,km.
Models with a value of the compactness parameter below $0.45$ can likely explode, while models $\ge 0.45$ undergo direct black hole formation \citep{O'Connor2011} (although this simple prescription has been challenged recently; see \citealt{Wang2022}).
To evaluate the compactness parameter, the stellar models are evolved beyond core He-exhaustion, except for the most massive models which are not expected to have a CCSN. 

For the SN energies, we use a linear fit to the numerical results in \citet{Schneider2021}, where the explosion energies appear correlated to the compactness parameter.
For a simplified estimate of the ejected mass, we assume that the innermost \SI{2}{M_{\sun}} of the stellar models are locked in the remnant, while the remaining mass is ejected. The ejected mass of hydrogen, helium and metals is calculated from the respective masses of these components outside the innermost \SI{2}{M_{\sun}}.

The compactness parameter is highly non-monotonic with stellar mass, resulting in the distribution of explosion energies shown in Fig.~\ref{fig:Stellar_SN}.
All single stellar models with masses around \SI{10}{M_{\sun}} explode, while some of the models around \SI{20}{M_{\sun}} do not. Which models explode varies with metallicity and initial rotation velocity and there is no simple relationship between explosion (or not) and the initial stellar parameters. 
Consistent with part of the literature, for an initial mass of \SI{25}{M_{\sun}} all models explode.
For binary systems, the lowest initial primary mass of a system where a CCSN takes place (either primary and secondary) is around \SI{16}{M_{\sun}}. In systems with an initial primary mass of \SI{16}{M_{\sun}} and \SI{25}{M_{\sun}}, all primaries and secondaries undergo a CCSN. For higher masses, CCSNe become rare for the primaries; however,  we find \citep[in broad agreement with][]{Schneider2021} that, in binary systems, higher mass stars also explode (up to \SI{60}{M_{\sun}}) compared with single stars. Nearly all secondaries explode where the initial primary mass is \SI{40}{M_{\sun}}; above that mass CCSNe only happen for a few models.

The SN energies span an order of magnitude, ranging from \SI{2.5e50}{erg} to \SI{2.5e51}{erg}, where more massive models overall tend to have higher energies. All models with a mass of \SI{8}{M_{\sun}} explode with nearly the same energy of around \SI{2.5e50}{erg}. No clear trends with initial rotation velocity for single stars or orbital period for binary systems are visible. 

By employing the compactness parameter to infer SN energies, the SN energy ejected by a stellar population varies with metallicity (because the compactness at core collapse depends in a non-linear way on many initial stellar properties, including metallicity).
As an example, a population of single non-rotating stars ejects at $Z=0.004$ an energy of 
\SI{4.1e49}{erg\,M_{\sun}^{-1}} 
from CCSNe, while the same population ejects at $Z=0.001$ only 
\SI{2.4e49}{erg\,M_{\sun}^{-1}} 
(weighted by a Salpeter initial mass function (IMF) normalised to the total mass in stars above
\SI{8}{M_{\sun}}; see Sect. \ref{sec:Results:overview_energy_flow} for a description of the adopted IMF weighting).
If instead one would employ the simple assumption that just all the single models (up to the highest mass of \SI{160}{M_{\sun}}) undergo a CCSN with the canonical energy of \SI{e51}{erg}, the population would provide 
\SI{5.0e49}{erg\,M_{\sun}^{-1}}, 
while  
for the whole grid (i.e. assuming every considered model undergoes a CCSN with \SI{e51}{erg} and a binary mass fraction of $70$ per cent), the population would provide 
\SI{4.7e49}{erg\,M_{\sun}^{-1}}, 
independent of metallicity.

Ejected masses increase with initial stellar mass and range from a few M$_\odot$ to around \SI{25}{M_{\sun}} 
and \SI{40}{M_{\sun}} 
for single and binary models, respectively. The few cases of extreme mass ejection of binary SN might originate from adopting the compactness parameter as the explosion criterion.
Overall, single models with a lower rotation velocity have higher SN ejecta masses due to the lower mass-loss rates over the course of their lifetime. In closer binary systems, less mass is ejected by the primary during the CCSN and more by the secondary compared to wider systems, due to the stronger mass transfers. However, the extent of these differences varies with metallicity. The metal fractions of the ejected masses also increase with initial mass. For single stars, the ejecta of fast rotating stars has higher metal fractions, up to where nearly all the ejected mass is found in metals for the most massive, H-depleted stars at the lowest metallicity. For binary systems, the CCSNe of primary stellar models have on average a higher metal fraction in their ejecta than the CCSNe of their secondaries.
In binary systems where both the primary and secondary undergo a CCSN, the metal fraction of the primary is higher than that of the secondary, due to the higher mass-loss rates by both Case A
mass-transfer 
\footnote{Mass-transfer where the primary is on the main sequence.}
as well as stellar wind. Low mass secondaries have the lowest metal fractions.
In the most extreme cases of stellar models undergoing SN explosions, up to nearly \SI{18}{M_{\sun}} (single) and \SI{30}{M_{\sun}} (binary) of metals would be ejected and then enrich the surrounding ISM. The progenitors of these cases are stripped of their outer layers already during their lifetime.

Overall, SN energy, mass and metal ejecta vary with metallicity. However, no strong trends with metallicity are visible (except a decrease of the mass ejecta with increasing metallicity for single stars). Binary systems give rise to CCSNe from progenitors with higher initial mass than single stars, leading to more energetic events. Moreover, binary star systems can potentially host a sequence of two successive SN events.

\section{Numerical methods}
\label{sec:Numerical methods}
\subsection{One-dimensional simulations}
We perform 1D, spherically symmetric simulations with the radiation-hydrodynamics code \textsc{Pion}, using the source term for evolving stellar winds described in \citet{Mackey2021}.
The stellar models described in Sect. \ref{sec:Stellar_models} provide the stellar mass, luminosity, effective temperature, mass-loss rate, rotation velocity, radius and element abundances
as a function of time.
These quantities are used to calculate (i) the ionising photon luminosity and mean energy per ionisation for the photoionisation calculation, and (ii) wind density, pressure and terminal velocity imposed in a boundary region around the origin.
The stars impact the surrounding ISM during their lifetime via kinetic winds injected into the innermost two cells, and via photoionisation energy injected throughout the H\,\textsc{ii} region, where most is injected at the Str\"omgren radius due to the high opacity.
The wind terminal velocity is calculated as described in Sect. \ref{sec:Stellar_models_life}.
The ionising photon luminosity, $Q_0$, is calculated by matching the temperature-dependent $Q_0$ data tabulated in \citet{DiazMiller1998} using a power-law fit below 33\,000\,K, scaled to the appropriate stellar radius.
At higher temperatures a blackbody provides a good estimate of $Q_0$.
The mean energy per ionisation is calculated assuming a blackbody spectral shape above 13.6\,eV.
The non-equilibrium photoionisation scheme was tested in \citet{Mackey2012} and benchmarked against analytic solutions and other codes in \citet{Bisbas2015}.

The radiative heating and cooling scheme was described in \citet{Mackey2013} and consists of: radiative heating by photoionisation, metal-line cooling assuming collisional ionisation equilibrium (CIE) with scaled solar abundances, Bremstrahlung in ionised gas, fine-structure line cooling of C$^+$, O in warm neutral gas, together with line cooling from CNO ions in photoionised gas, similar to the model of \citet{Henney2009}.

For this project we modified the 2D/3D nested grid algorithms \citep{Mackey2021} to also work for 1D spherical coordinates, focused on the origin.
The 1D nested grid is a simple reduction of the multidimensional algorithm with the addition of appropriate geometric source terms. 
We follow the description in \citet{Mackey2021} of prolongation and restriction \citep{Toth2002} at the coarse/fine level boundaries, and ensure consistency between levels using the flux correction algorithm of \citet{Berger1989}. 

The coarsest level 0 covers the full domain of size $\mathcal{L}$ with radial coordinate $r\in[0,\mathcal{L}]$; level 1 covers $r\in[0,\mathcal{L}/2]$, level $j$ covers $r\in[0,\mathcal{L}/2^j]$, down to the finest level.
The coordinate singularity at $r=0$ is treated with reflecting boundary conditions (the surface areas of this boundary is anyway zero so the flux through $r=0$ is always zero), and in the pre-SN phase the stellar-wind boundary condition is imposed on the first two cells adjacent to $r=0$.
The outer boundary of each level $j$ is obtained by interpolating data from the appropriate cells of level $j-1$ and these are updated every timestep.
After each step of level $j$, averaged states are sent to level $j-1$ to overwrite cell data in the subdomain $r\in[0,\mathcal{L}/2^j]$ of level $j-1$.
An integration scheme is used that is accurate to second order in time and space \citep{Falle1998}, using a TVD linear interpolation and minmod slope limiter.

\subsection{Initial conditions and simulation parameters}
The simulation suite is performed in a computational domain of 
$\approx [0,66]$ pc, which ensures that the feedback bubble does not leave the domain during the star's lifetime.
We use six levels of static mesh-refinement with 256 grid cells per level and a factor of 2 refinement with each level as described above, resulting in a cell size of $0.0081$ pc near the origin to $0.26$ pc on the coarsest level at $r>\mathcal{L}/2$. 
In cases where the free-streaming region of the wind at $10$ per cent of the (primary) stellar lifetime is not resolved with at least 5 grid cells (happening only for the lowest-mass stellar models), we add additional refinement layers in the centre until the criterion is met.

We use a uniform ISM with four different densities: 
\num{2.2}, \num{4.4}, \num{11}, and \SI{22e-22}{\gram\per\cubic\centi\meter} 
(corresponding to a hydrogen number density of around \num{100}, \num{200}, \num{500}, and \SI{1000}{cm^{-3}}).
These densities are selected to approximate the range of SF densities in zoom-in simulations of galaxies.
While in general the ISM is actually turbulent and structured, this cannot be captured in spherically symmetric calculations and a uniform ISM is as realistic as any other feasible configuration.
Multidimensional simulations would be preferable, but at the cost of orders of magnitude more computation and storage.
Even if these are possible, 1D simulations provide a valuable point of reference to investigate the importance of multidimensional effects such as dynamical instability and turbulent mixing.

The temperature of the ambient medium is initialised at $50$ K, which is within a factor of two of the equilibrium temperature for the different setups. 
A temperature floor of \SI{10}{\kelvin} is set.
Initial ambient metallicities are set to the initial surface values of the stellar models.
Tracers for H, He, C, N, O, Z, and H+ are passively advected, where in Z the remaining metals are stored.
Non-equilibrium ionisation of H$^+$ is calculated including photoionisation, collisional ionisation and radiative recombination, using a backward-difference implicit method.
Ionising photon radiation is calculated by raytracing outwards from the star at the origin to track the attenuation.
Outputs are generated every \SI{e11}{\second} and the simulations are run until the end of the stellar lifetime.
In total, the \textsc{Pion} simulation grid consists of $1960$ runs of single stars and $560$ runs of binary systems.

To simulate the subsequent SN, we use the \textsc{Pion} output at the end of the stellar lifetime. We overwrite the innermost 10 cells, using a 2-component profile for the density and velocity
following \citet{Whalen2008}, where the two parameters of the model are found iteratively. 
For the SN of the primary in a binary system, the SN injection is shifted by two cells to avoid conflicts with the wind injection of the secondary.
The hydrogen and helium fractions are injected into their corresponding tracer fields, while the metals of the SN are traced with a separate field.
Outputs are generated in multiples of the time step, where the early evolution is sampled with a higher frequency.
The SN simulations are evolved until \SI{5}{Myr}, or longer until the shell of material swept up by the SN reaches \SI{20}{pc}.
For the binary grid, the simulation is restarted from the last output of the primary lifetime to simulate the remaining secondary lifetime, adding, if required, a primary SN. In case the secondary has a CCSN, its evolution is simulated as in the single star case. In cases with a primary SN and no secondary SN, we ensure that the simulation is run for at least \SI{5}{Myr} after the primary SN.

\section{Results}
\label{sec:Results}
\subsection{Evolution of selected models}
First, we look at selected models in more detail: These models are single stellar models with $Z=0.004$ and without stellar rotation, in a uniform medium of $\SI{100}{cm^{-3}}$. 

\subsubsection{Evolution of a 32 M$_\odot$ star}
\label{sec:Results:evolution_32}
\begin{figure}
    \centering
    \includegraphics[width=1\columnwidth]{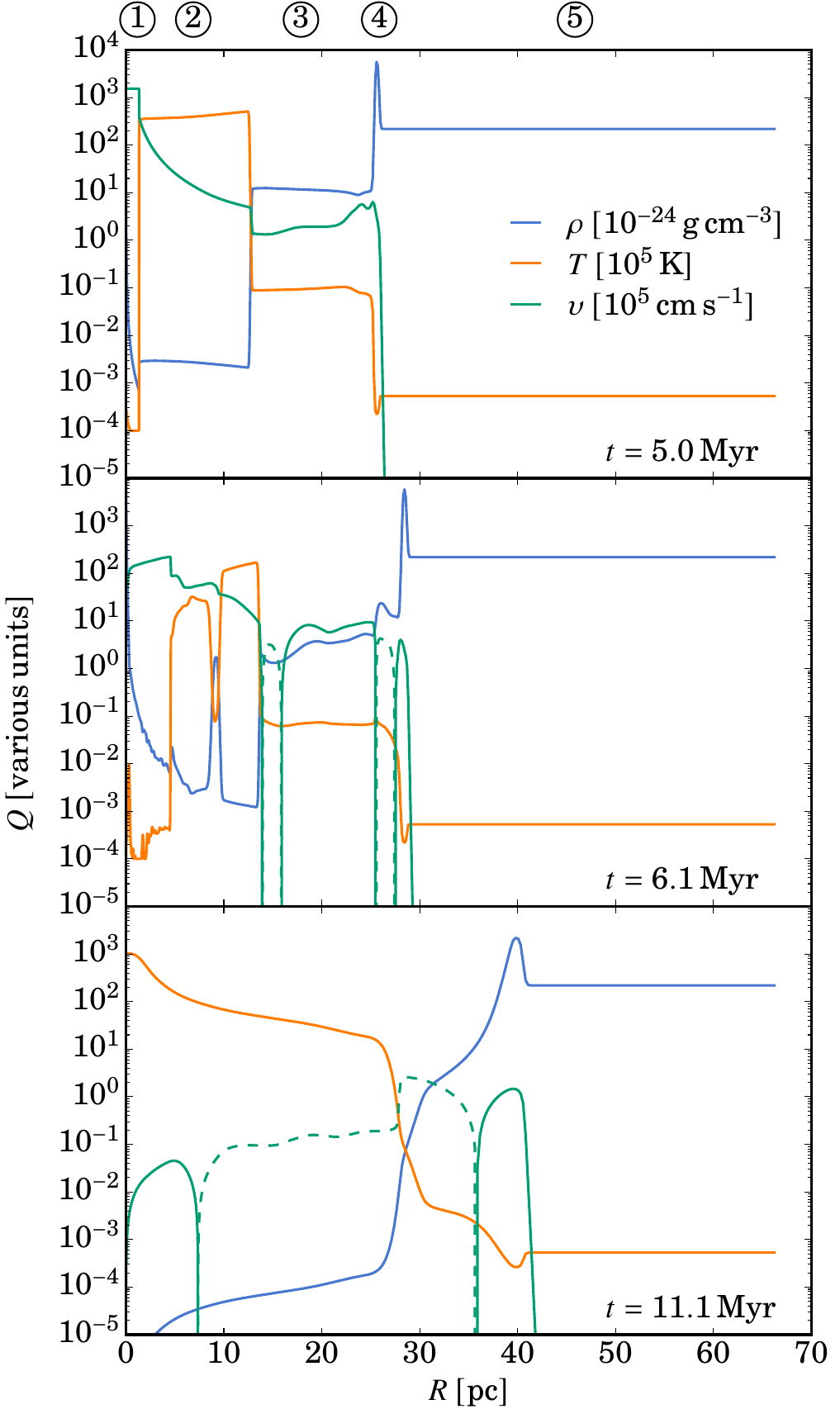}
    \caption{Density, temperature, and velocity profiles of a \textsc{Pion} simulation with a single star model of $Z=0.004$, $\upsilon_\textrm{ini}=\SI{0}{km\,s^{-1}}$, and $\log_{10}(M/\mathrm{M}_\odot)=1.5$ in a uniform medium of $n=\SI{100}{cm^{-3}}$ at \SI{5}{Myr} (top), at the end of the stellar lifetime (middle), and \SI{5}{Myr} after the subsequent SN (bottom). Dashed lines indicate negative velocities. The circled numbers at the top indicate the five different regions found in the density profile due to pre-SN feedback.}
    \label{fig:Results_Profile_32Msun}
\end{figure}

We show the density, temperature and velocity profiles of one \textsc{Pion} simulation to examine the general evolution in the circumstellar medium around a massive star.
In the top panel of Fig.~\ref{fig:Results_Profile_32Msun}, five regions can be seen, which are indicated by the encircled numbers above the top panel. These are from left to right: 1: free-expansion of the stellar wind, 2: hot shocked wind material, 3: H\,\textsc{ii} region, 4: cold swept-up medium, and 5: unperturbed surrounding medium.  
In the middle panel, which shows the last output during the star's life, shells of wind ejecta can be seen as locally overdense and cooler regions (in Fig.~\ref{fig:Results_Profile_32Msun} at around $10$ pc). At the end of the stellar lifetime, the feedback bubble is around \SI{28}{pc}. Therefore, the pre-SN feedback displaced already around \SI{3e5}{M_{\sun}} of circumstellar material.
The bottom panel shows the profile at the end of the SN simulation which was evolved for \SI{5}{Myr}, containing a hot, low-density medium
surrounded by
a shell of cold swept-up gas. The overdense shells and other structures found at the end of the star's life are swept by the SN ejecta. In the \SI{5}{Myr} after the SNe, the bubble expands to around \SI{40}{pc}, displacing around \SI{9e5}{M_{\sun}}.
In Appendix~\ref{sec:Appendix_A}, we show profiles for the evolution of a binary system with two SNe for comparison.

\begin{figure}
    \centering
    \includegraphics[width=1\columnwidth]{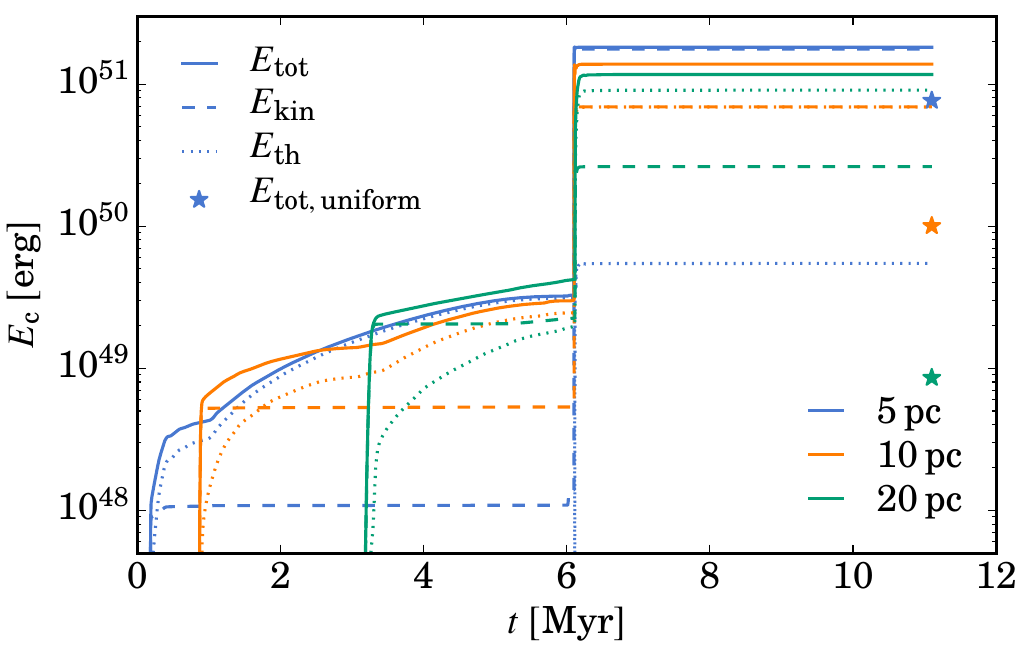}
    \caption{Cumulative energy flow through $r_\textrm{s}=5$, $10,$ and $20$ pc 
    for a single star model of $Z=0.004$, $\upsilon_\textrm{ini}=\SI{0}{km\,s^{-1}}$, and $\log_{10}(M/\mathrm{M}_\odot)=1.5$ in a uniform medium of $n=\SI{100}{cm^{-3}}$. The solid line shows the total energy, while the dashed and dotted lines show the kinetic and thermal energy component, respectively. A SN explosion occurs around \SI{6.1}{Myr}. The star symbols (labelled `uniform') show a comparison of the total energy with respect to a SN run in a uniform medium without pre-SN feedback beforehand. }
    \label{fig:Results_Flow_32Msun}
\end{figure}

We shall now discuss the energy flow through surfaces of spheres at different radii, and equate the energy that is available for feedback in cell sizes above a few pc. 
These radii are denoted $r_\textrm{s}$ and we choose $r_\textrm{s}\in[5,10,20]$ pc for this work because they cover the full range of behaviours (i.e. barely any dissipation to strong dissipation)
at the gas densities considered.
The kinetic and thermal energy densities, $u_\textrm{kin}$ and $u_\textrm{th}$, at a radius of $r_\textrm{s}$ and time $t$ since a star's birth are 
\begin{align}
u_\textrm{kin}(r_\textrm{s},t)&=0.5\,\rho(r_\textrm{s},t)\,\upsilon(r_\textrm{s},t)^2 \,, \\ 
u_\textrm{th}(r_\textrm{s},t)&=\frac{p(r_\textrm{s},t)}{(\gamma-1)} \mathrm{,}
\end{align}
with $\rho$ the (mass) density, $\upsilon$ the 1D-velocity, $p$ the pressure, and $\gamma=\frac{5}{3}$.
The energy flux through the spherical surface, $e_\textrm{i}(r_\textrm{s},t)$ is then
\begin{equation}
    e_\textrm{i}(r_\textrm{s},t)=u_\textrm{i}(r_\textrm{s},t)\,\upsilon(r_\textrm{s},t)\,4\,\pi\,r_\textrm{s}^2 \;,
\end{equation}
and the cumulative energy flow through this sphere is the discretised time integral of the flux:
\begin{equation}
    E_\textrm{i,\,c}(r_\textrm{s},t)=\sum_{t_\textrm{j}\leq t} e_\textrm{i}(r_\textrm{s},t_\textrm{j})\,\Delta t  \;,
\end{equation}
with $i$
indicating
the kinetic or thermal energy density and $\Delta t$ the time interval with respect to the last output.
We only consider radii of $r_\textrm{s}=20$ pc and below, as we find (see e.g. Sect. \ref{sec:Results:overview_energy_flow}) that
nearly all energy is dissipated already for $20$ pc in the cases of higher densities. 
A caveat in the flow calculations is introduced by the photoionisation energy. It is injected mostly at the Str\"omgren radius; if that radius is larger than $r_\textrm{s}$, the photoionisation energy is not accounted for.

Figure \ref{fig:Results_Flow_32Msun} shows the cumulative energy flow (total, and split into the kinetic and thermal component) through $r_\textrm{s}=5$, $10$ and $20$ pc of this model. 
In the pre-SN flow, before the end of the stellar lifetime at around \SI{6.1}{Myr}, we can see that the kinetic energy only flows through each sphere at one instance in time when the cold swept-up mass crosses that radius, while the thermal energy flow is more continuous. However, also the thermal energy shows a noticeable rise at two times.
The first occurs when the swept-up mass crosses $r_\textrm{s}$, 
(where also the kinetic energy increases strongly)
and the second rise occurs
when the first shocked wind material reaches $r_\textrm{s}$.
Higher radii are reached by the swept-up mass at a later time, for example, it takes around \SI{1}{Myr} until the first feedback energy reaches $10$ pc. The kinetic energy is higher at higher radii at the end of the stellar lifetime (i.e. before around \SI{6.1}{Myr}), 
while the total energy shows a more complicated behaviour with radius due to the interplay of energy dissipation and the injection of photoionisation energy at radii larger than $r_\textrm{s}$.
The fraction of kinetic energy to the total energy at the end of the stellar life increases from around $4$
per cent at $5$ pc to around $50$
per cent at $20$ pc. The shocked wind material never reaches $20$ pc, so we see there mainly the contribution of the photoionisation energy to the total energy. 

The subsequent SN explosion of the stellar model at around \SI{6.1}{Myr} leads to a sharp rise in the cumulative energy flow, dominating the total energy budget at all radii. Due to the low density inside the feedback bubble at the end of the stellar lifetime, the energy from the SN explosion can rapidly reach $20$ pc without conspicuous radiative energy losses. At $5$ pc, nearly all the SN energy is kinetic, while at $20$ pc part of the energy has been converted to thermal energy through shock heating.

Although SN energy is the predominant factor in the overall energy budget, the pre-SN feedback occurring during the star's lifespan is crucial. We present a comparison involving the same SN, but situated in a uniform medium unaffected by the star's pre-SN feedback, in Fig. \ref{fig:Results_Flow_32Msun}. This is depicted with symbols positioned at the conclusion of the fiducial simulation.
For all radii, the simulation without pre-SN feedback has a lower total energy. The differences increase with radius. The case without pre-SN feedback has dissipated the energy by one and two orders of magnitude for $10$ and $20$ pc, respectively. On the contrary, even at $20$ pc, our simulation with pre-SN feedback has retained
most of its energy. This results in a difference of two orders of magnitude in the energy at $20$ pc. Further comparisons with the case without pre-SN feedback are shown in the Sect. \ref{sec:Results:without_preSN_feedback}.

\subsubsection{Density profiles of one sequence of stellar models}
\begin{figure}
    \centering
    \includegraphics[width=1\columnwidth]{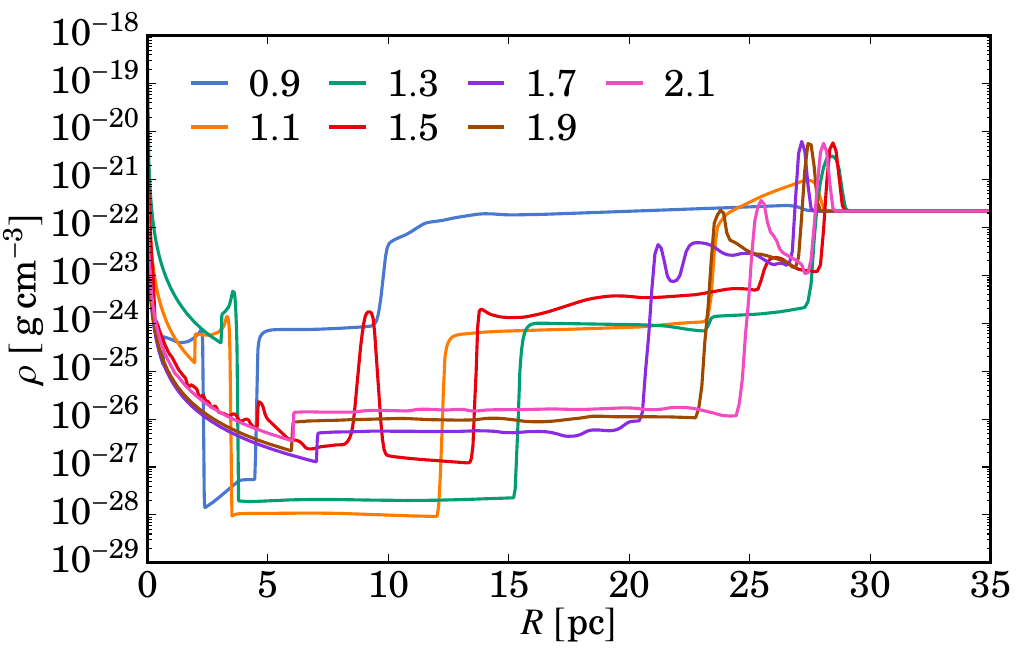}
    \caption{Density profiles of our simulations with single star models of $Z=0.004$, $\upsilon_\textrm{ini}=\SI{0}{km\,s^{-1}}$, in a uniform medium of $n=\SI{100}{cm^{-3}}$ at the end of the star's lifetime. For clarity, we show only the profiles of very second model of this sequence of stellar models. The labels indicate the logarithm of the initial masses in \si{M_{\sun}}.}
    \label{fig:Results_profiles}
\end{figure}
\begin{figure}
    \centering
    \includegraphics[width=1\columnwidth]{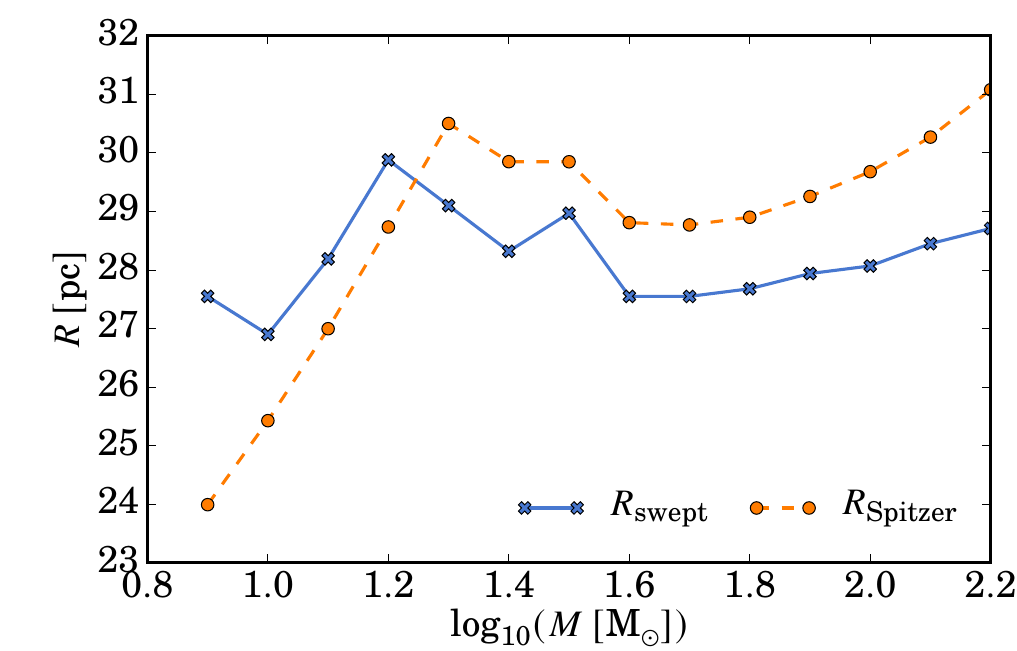}
    \caption{Comparison of the radius of the outer edge of the shell of swept-up material to an estimate from the Spitzer radius. The radius is calculated from the single star models of $Z=0.004$, $\upsilon_\textrm{ini}=\SI{0}{km\,s^{-1}}$, in a uniform medium of $n=\SI{100}{cm^{-3}}$ at the end of the star's lifetime. }
    \label{fig:Results_Spitzer_comparison}
\end{figure}
Figure~\ref{fig:Results_profiles} shows the density profiles of the half of the stellar models at $Z=0.004$, $\upsilon_\textrm{ini}=\SI{0}{km\,s^{-1}}$, in a uniform medium of $n=\SI{100}{cm^{-3}}$ at the end of the stellar lifetime. 
It is remarkable that the size of the feedback bubble (i.e. the location of the peak of the swept-up mass at around \SI{27}{pc}) is similar across nearly the whole mass range.
To illustrate the origin of the similar size, we compare the size of the feedback bubble (measured by the outer edge of the swept-up shell)
to the relation for the radial evolution of an H\,\textsc{ii} region \citep{Spitzer1978} in Fig.~\ref{fig:Results_Spitzer_comparison}.
For estimating the Spitzer radius, 
we use the properties (luminosity, temperature, lifetime) of the simulated stars and the temperature of the H\,\textsc{ii} region, but we neglect any time evolution of these.
The H\,\textsc{ii} region radius calculated from the Spitzer solution is also nearly independent from the stellar mass for the highest mass models. The small dependence that is visible is similar in both and the radii agree reasonable well considering the approximations that were made.
The reason for the nearly mass independence is that the lower luminosities of ionising photons for smaller stellar masses is compensated by longer lifetimes (in that stellar mass range).
We note that the H\,\textsc{ii} regions of the lower mass models have reached pressure equilibrium with their surroundings, and so the neutral shell of swept-up material becomes geometically thick, invalidating the assumptions in the Spitzer solution \citep[see also][]{Williams2018}. For higher mass stars this is not the case and a thin shell of swept-up gas bounds the H\,\textsc{ii} region (see Fig.~\ref{fig:Results_profiles}).

\subsubsection{Evolution of a 100 M$_\odot$ star}

The energy flows for our single-star models with a metallicity of $Z=0.004$ and no stellar rotation are depicted in Fig.~\ref{fig:Results_Flow_z004_r000_single}. The model with \SI{100}{M_{\sun}} shows that the swept-up mass reaches the different $r_\textrm{s}$ earlier compared to the \SI{32}{M_{\sun}} model. The progression of energy flow in both models is similar, yet with higher values for the \SI{100}{M_{\sun}}, up to about $3$ million years, which marks the onset of the WR phase. Following this, there is a pronounced increase in cumulative energy, with shorter intervals compared to the initial gradient. At a radius of $r_\textrm{s} = 5$ pc, the energy composition before the WR phase is predominantly thermal as the termination shock lies within this radius. After the WR phase, the kinetic energy becomes dominant as the termination shock moves beyond this radius. Conversely, at $r_\textrm{s} =20$ pc, the trend is opposite: kinetic energy is prominent before the WR phase, but post-phase, thermal energy takes precedence. This indicates that the kinetic energy from the WR wind is transformed into thermal energy as it travels outward. The shocked wind material extends to $20$ pc, with the WR wind pushing a shell of early wind material beyond even $20$ pc. Notably, in this model, the total energy diminishes as the radius increases, and it does not undergo a SN explosion.

\subsubsection{Evolution of a 10 M$_\odot$ star}

As opposed to the \SI{32}{M_{\sun}} model, Fig.~\ref{fig:Results_Flow_z004_r000_single} shows that the cumulative thermal energy flow driven by a \SI{10}{M_{\sun}} star dominates over the kinetic energy flow at all radii during most of the star's lifetime. Only when the swept-up mass reaches $r_\textrm{s}$ is the kinetic energy important. This time is later than in the \SI{32}{M_{\sun}} case. The total energy is increasing with radius in the pre-SN phase and exceeds the injection of the stellar wind for $10$ and $20$ pc as the energy budget is completely dominated by photoionisation. The photoionisation energy couples mostly at the Strömgren radius, and therefore the energy is not yet injected at the smaller radii. 
The subsequent SN dominates the total energy flow by more than one order of magnitude and is decreasing with distance. As the stellar model created a smaller wind cavity compared to the \SI{32}{M_{\sun}} model, the SN energy remains mostly in the kinetic component only out to $\sim5$ pc. At $10$ pc, most of SN energy has been converted to thermal energy. Further out at $20$ pc, the total energy decreases by over one order of magnitude due to thermal dissipation, only a small kinetic energy flow remains.

\subsection{Overview of the cumulative energy flow}
\label{sec:Results:overview_energy_flow}
\begin{figure*}
    \centering
    \includegraphics[width=1\textwidth]{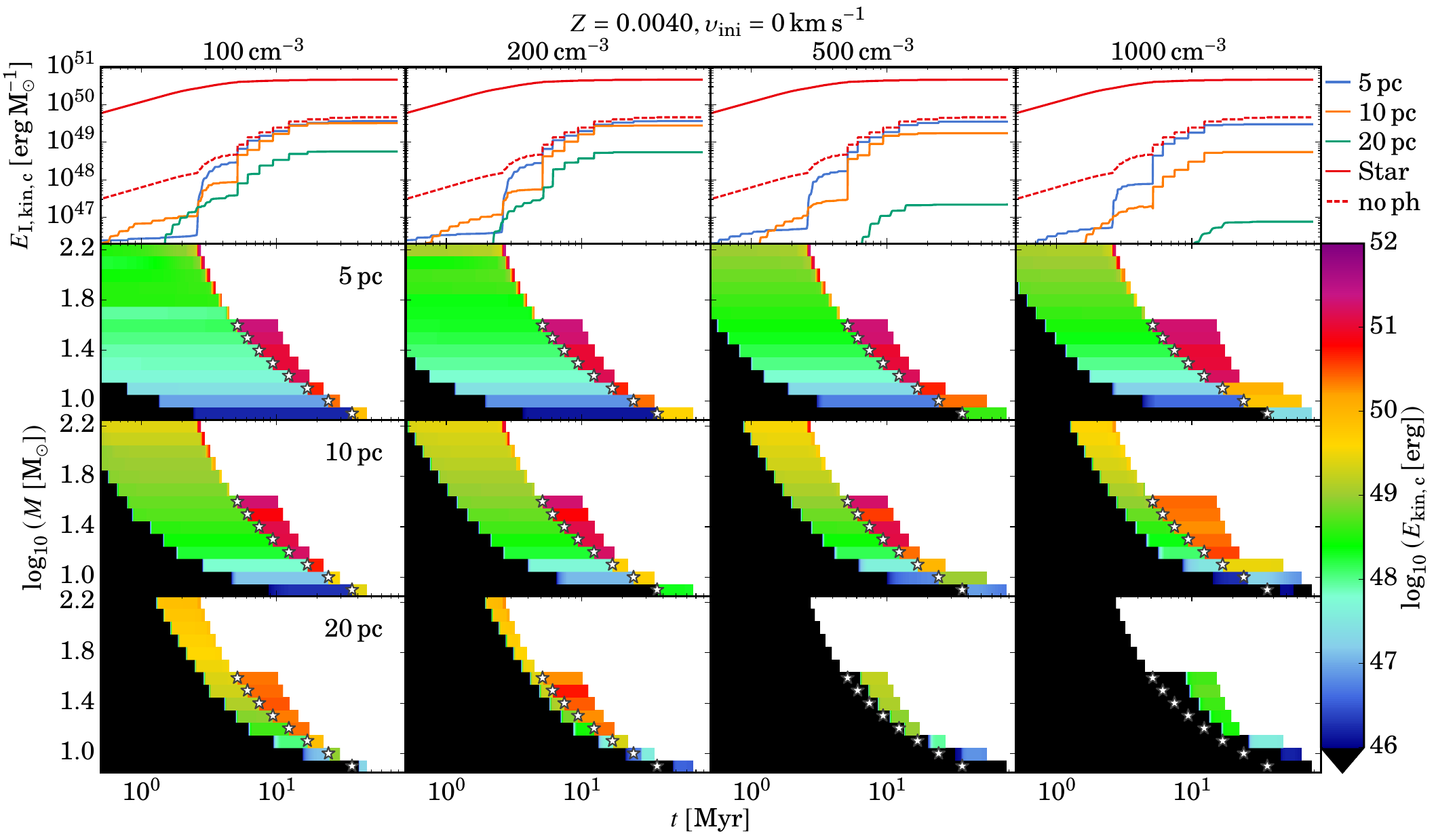}
    \includegraphics[width=1\textwidth]{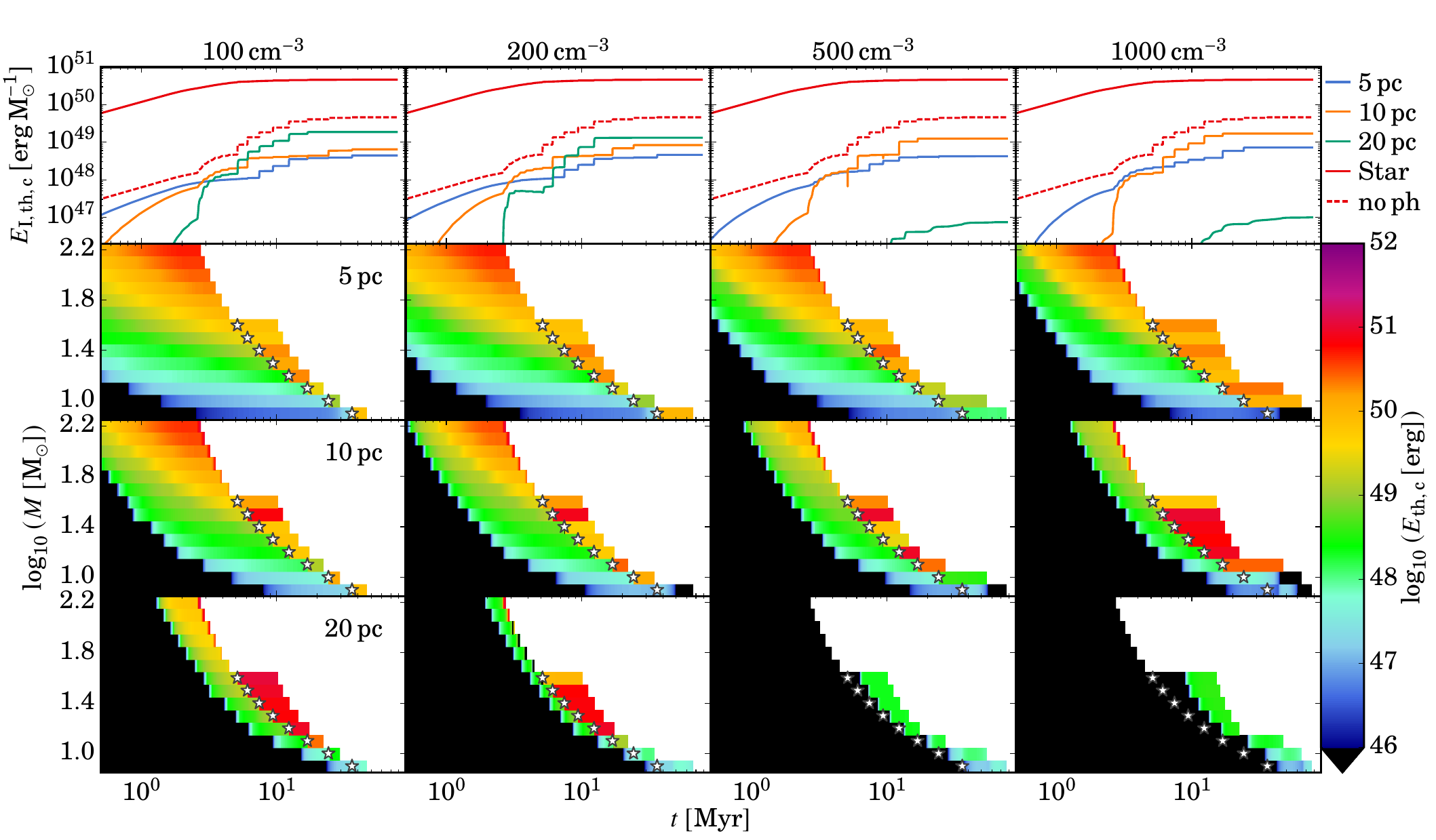}
    \caption{Cumulative kinetic (top) and thermal (bottom) feedback energy for different stellar initial mass as a function of time for the single star models with $Z=0.004$ and no initial rotation velocity is shown with coloured bars for the whole time of the \textsc{Pion} simulation. The different columns show different initial ambient densities, while the lower three rows vary $r_\textrm{s}$. The star symbol indicates the time of the SN, while black regions visualise outflowing energies below \SI{e46}{erg} or inflows. The upper row shows IMF-weighted results (normalised to the total mass in stars above \SI{8}{M_{\sun}}). Additionally, the sum of the energy inputs from SN, stellar winds and an estimate of the photoionisation energy is shown (labelled `Star'), and the energy input excluding the photionisation energy (labelled `no ph').}
    \label{fig:Results_Flow_z004_r000_single}
\end{figure*}

In addition to the flows of the different stellar models, Fig.~\ref{fig:Results_Flow_z004_r000_single} shows IMF-weighted values using a Salpeter IMF for our models between $10^{0.9}$ M$_\odot$ for single stars, or $10$ M$_\odot$ for the primary star mass in binary systems, to $10^{2.2}$ M$_\odot$. 
We note that we normalise for a population of only massive stars with masses between \SI{8}{M_{\sun}} and $10^{2.2}$ M$_\odot$ to avoid the more uncertain low mass end of the IMF. 
For binary systems, we apply the IMF to the primary star. 
When including the SN runs in the IMF, we neglect the evolution after the start of inflows characterised by a decrease in the energy.

We compare the IMF-weighted cumulative flows with the time-integrated input energy from the stellar wind, SN and photoionisation energy. The photoionisation energy is estimated by integrating the Planck spectrum from \SI{13.6}{eV} to \SI{100}{eV}, while subtracting \SI{13.6}{eV} from each photon, which is the maximum power available to drive feedback. This assumes that most ionising photons ionise a H atom and heat the gas with the remaining photon energy, and neglects radiative losses.

\subsubsection{Radial distance evolution of the cumulative energy flow}
\begin{figure}
    \centering
    \includegraphics[width=1\columnwidth]{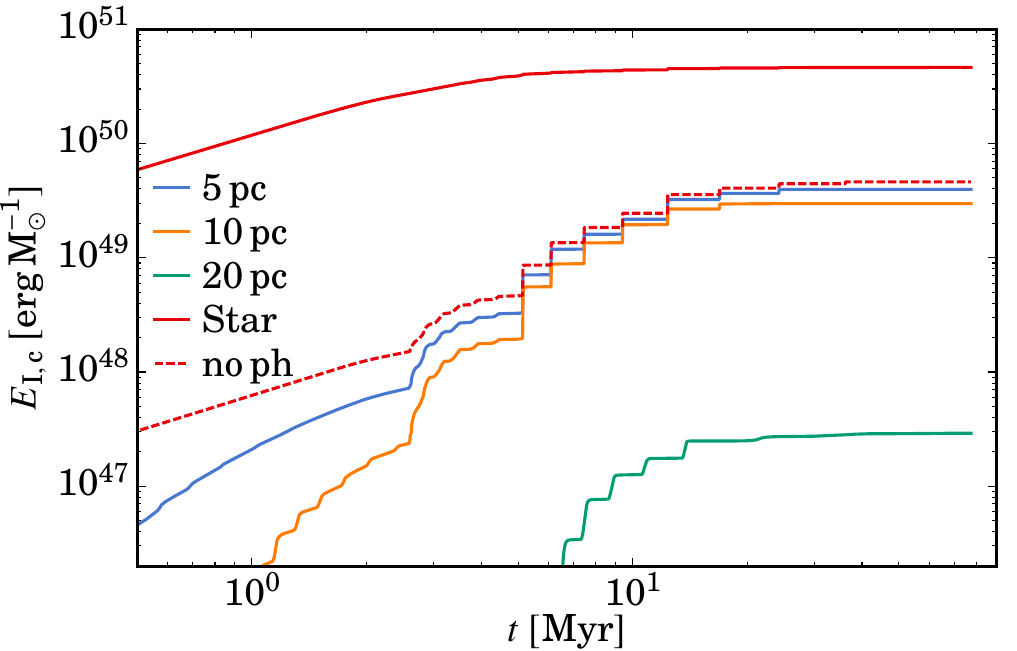}
    \caption{IMF-weighted (normalised to the total mass in stars above \SI{8}{M_{\sun}}) cumulative energy flow for the single star models with metallicity $Z=0.004$ and no initial rotation with a background density of \SI{500}{cm^{-3}}. Additionally, the sum of the energy inputs from SN, stellar winds and an estimate of the photoionisation energy is shown (labelled `Star'), and the energy input excluding the photionisation energy (labelled `no ph').
    }
    \label{fig:Results_Flow_z0001_r000_d0500_single}
\end{figure}

First we look at single star models with $Z=0.004$ and no initial rotation velocity.
We start with the total IMF-weighted cumulative flows of the density \SI{500}{cm^{-3}}, visible in Fig. \ref{fig:Results_Flow_z0001_r000_d0500_single}.
The energy reaching $r_\textrm{s}$ decreases with distance, resulting in orders of magnitude lower energy at $20$ pc. In addition, the energy crosses more distant values of $r_\textrm{s}$ later. This means that 
the first time the energy arrives at a large distance is a few million years after the birth of the stellar population.  
As for the separate models, the SN energy also dominates the IMF-weighted energy budget over the energy crossing $r_\textrm{s}$ during the stellar lifetime. 

Compared to the total energy input of winds, photoionisation, and SN, the energy flows from 1D simulations are significantly lower (see the red line in Fig.~\ref{fig:Results_Flow_z0001_r000_d0500_single}). However, the energy is similar to the injection of only wind and SN energy in both total energy and its time evolution, when evaluated at $r_\textrm{s}=5$ pc. This highlights the low coupling efficiency of the photoioninzation energy. For the smallest radius energy losses are negligible (especially at lower densities).
At early times (below around \SI{2}{Myr}), the energy flows are also lower with respect to only wind and SN injection. This highlights that main sequence stellar winds are not efficient, while later phases and SNe are. 

\subsubsection{Overview of the cumulative energy flow at fixed metallicity}
We compare the different background densities and split the energy into the kinetic and thermal components (see Fig. \ref{fig:Results_Flow_z004_r000_single}).
Before the onset of the first SN, corresponding to the early times in Fig. \ref{fig:Results_Flow_z004_r000_single}, 
the late evolutionary phases of the most massive stars are the biggest contributors to the kinetic energy. This can be seen from the sharp increase in the IMF-weighted kinetic energy at the time when they enter the post-main-sequence phase. Overall, the cumulative IMF-weighted kinetic energy flow before SNe can be well approximated by a single injection due to the steep increase caused by the post-main sequence phase. 
Similar to bigger values of $r_\textrm{s}$, higher background densities reduce the kinetic energy reaching $r_\textrm{s}$ as both higher densities and larger distances increase the mass within $r_\textrm{s}$.
The kinetic energy (including SNe) at $5$ pc is nearly independent of density, as the pre-SN feedback created a lower density feedback bubble beforehand, and therefore the central densities are independent of the initial densities. 
Where the IMF-weighted kinetic energy decreases strongly with increasing measurement radius, $r_\textrm{s}$, it indicates that the SN becomes radiative in between these radii.
As expected, this radius decreases with increasing density.

Similar to the IMF-weighted kinetic energy, the IMF-weighted thermal energy before the first SNe is dominated by the feedback from the most massive models.
For high background densities or large radii, the late phases of the most massive models become more important for the total thermal energy because part of the wind energy is converted from kinetic to thermal through shock heating. 
At low densities and large radii (see lower left and IMF-weighted subpanels in Fig. \ref{fig:Results_Flow_z004_r000_single}), the total energy is dominated by thermal energy from the late stages of the massive stars. 
The thermal energy flowing through $5$ pc does not vary strongly with density.

The IMF-weighted thermal to kinetic energy ratio changes with measurement radius.
The intermediate mass stars (\SI{25}{M_{\sun}} to \SI{40}{M_{\sun}}) are the strongest contributors to the total energy as they are the `best of both worlds'. They have SNe like stars around \SI{10}{M_{\sun}}, but the SN energy is higher and their winds are stronger. While the most massive stars around \SI{100}{M_{\sun}} have even stronger winds, they do not have SNe. This, in combination with the IMF, makes them important for the total energy feedback from a stellar population.

\begin{figure*}[h!]
    \centering
    \includegraphics[width=1\textwidth]{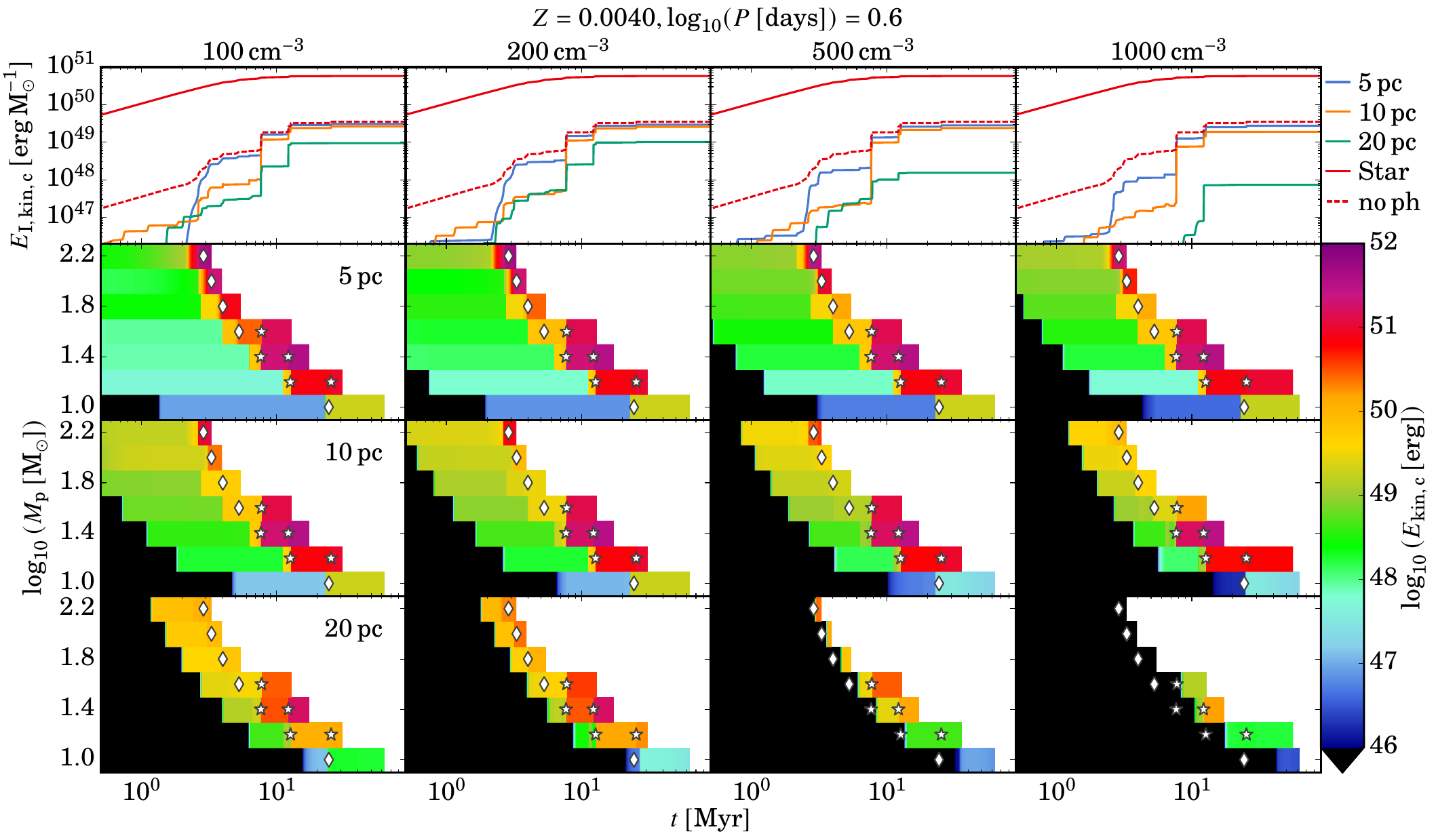}
    \includegraphics[width=1\textwidth]{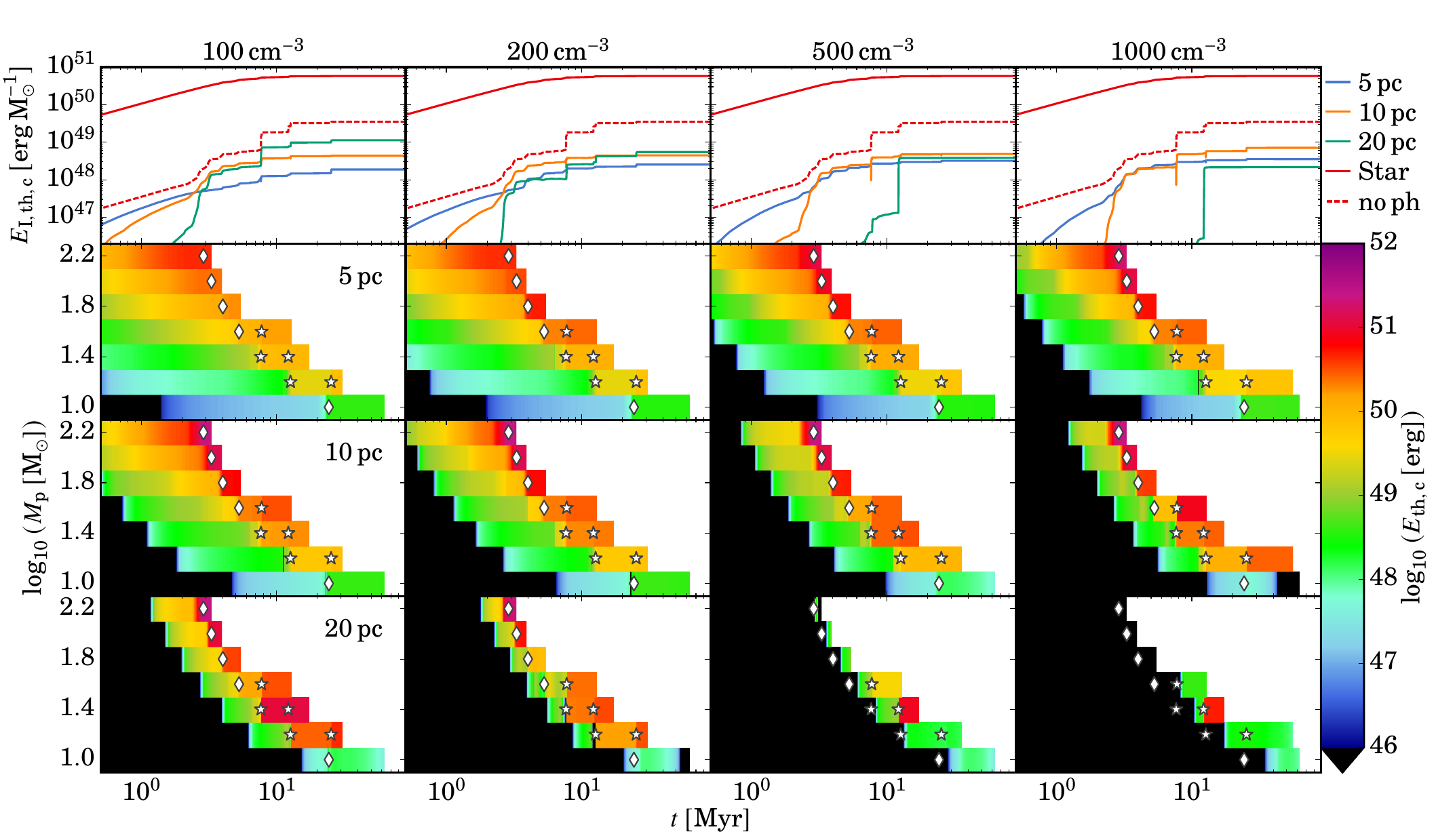}
    \caption{Same as Fig. \ref{fig:Results_Flow_z004_r000_single}, but for binary systems with $Z=0.004$ and $\log_{10}(P\mathrm{\,[days]})=0.6$. The star symbol indicates the time of a SN (primary or secondary), while the diamond symbol indicates the end of the primary lifetime without a SN.}
    \label{fig:Results_Flow_z004_p0.600_binary}
\end{figure*}
Figure \ref{fig:Results_Flow_z004_p0.600_binary} is the same as Fig. \ref{fig:Results_Flow_z004_r000_single} except that we show results for the grid of binary systems with the shortest period. Trends with density and radius are similar for the binary systems as for the single models. Also the comparison of the input energy with the measured energy flow is similar. Some results, however, are specific for the binary systems.
In the most massive binary systems, mainly the primary star contributes to the kinetic and thermal energy flow. Some models experience a phase of strong mass loss during mass-transfer events, which can lead to the formation of circumstellar shells. During the time when only the secondary star lives, there is barely any energy increase for most models (without SN). 
Compared to the single models, for binaries there is also the special case of runs with two subsequent SN when both primary and secondary undergo a CCSN. In these cases, the secondary SN evolves in a cavity formed by the primary SN, which reduces energy dissipation. This is especially important for the case of the highest density, where the SN of the secondary star of the $10^{1.4}$ M$_\odot$ primary retains nearly \SI{e51}{erg} even until \SI{20}{pc}, while the first SN in the system lost most of its energy. However, this is not the case for the secondary SN of the system with a $10^{1.2}$ M$_\odot$ primary due to the lower feedback.

Overall, when considering only the evolution during the stellar lifetime, we can say that
the hot bubble is resolved with $5$ pc. However, with $10$ pc it is not very well resolved, and with $20$ pc not at all.
On large scales these results favour a kinetic energy injection method, because the hot bubble is not resolved.
This is in agreement with recipes of feedback that do not inject thermal energy as it is being radiated away, but instead kinetic energy \citep[e.g.][]{DallaVecchia2008}.
The plots indicate that the main sequence of the stellar model is not a main energy contributor. The thermal energy is dominated by photoionisation, the kinetic energy by the post-main-sequence phase during the star's lifetime.
Far away from the star, a delay of the arrival of significant energy of more than \SI{1}{Myr} occurs.
The kinetic to total energy ratio changes with time, density, radius and stellar track.

\subsubsection{Metallicity comparison of the cumulative energy flow}
\begin{figure*}
    \centering
    \includegraphics[width=1\textwidth]{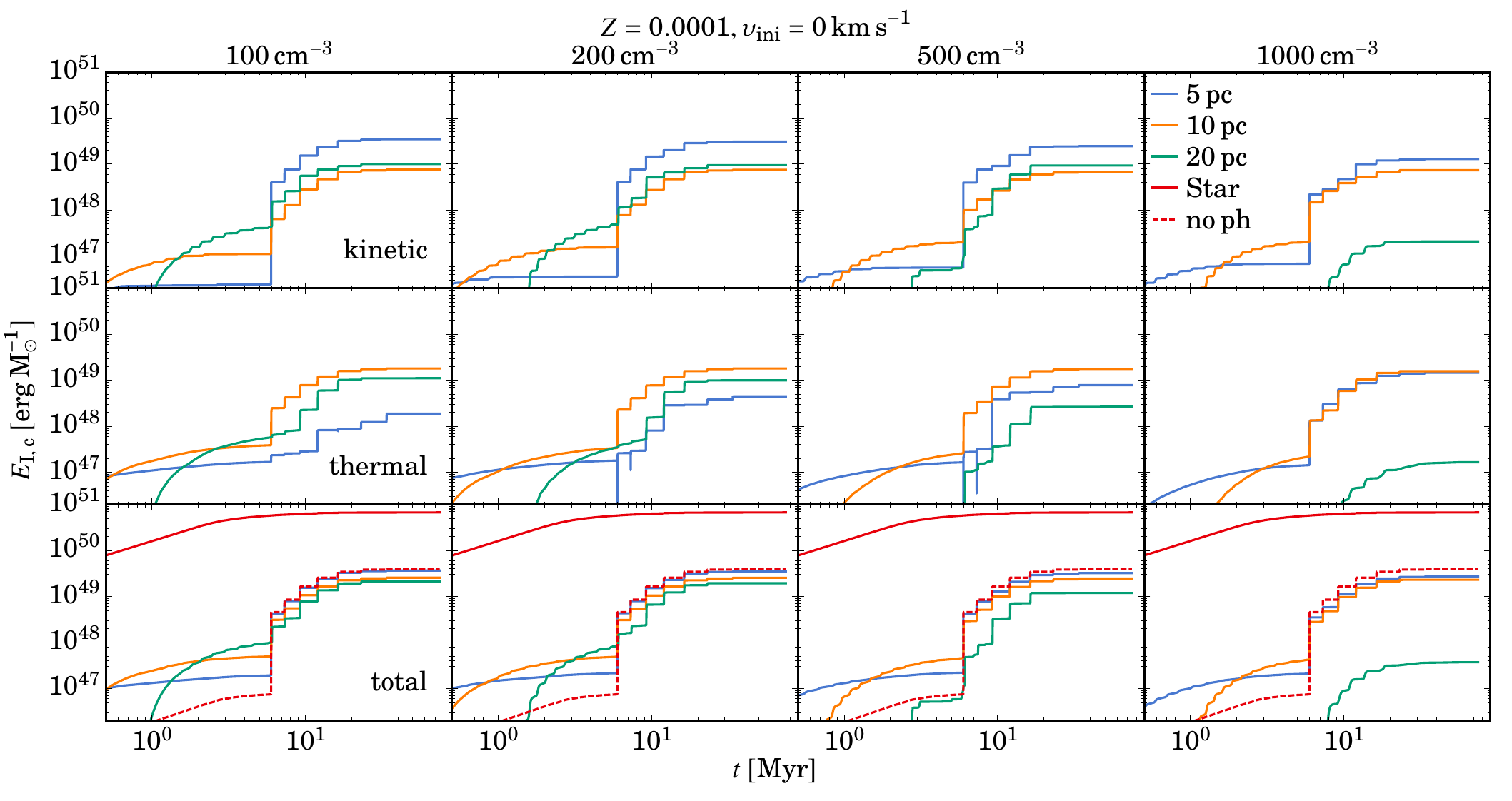}
    \caption{
    IMF-weighted (normalised to the total mass in stars above \SI{8}{M_{\sun}}) cumulative energy flow for the single star models with metallicity $Z=0.0001$ and no initial rotation. From top to bottom, the rows in each panel show kinetic, thermal, and total energy flow, while the columns show different initial ambient densities. Additionally, the sum of the energy inputs from SN, stellar winds and an estimate of the photoionisation energy is shown (labelled `Star'), and the energy input excluding the photionisation energy (labelled `no ph').
    }
    \label{fig:Results_Flow_z0001_r000_single}
\end{figure*}

At lowest metallicity (see Fig. \ref{fig:Results_Flow_z0001_r000_single}), the SN dominates the kinetic energy flow by orders of magnitude over the pre-SN feedback (stellar winds and photoionisation). 
The thermal energy is similar for all densities for $5$ and $10$ pc, indicating that the SN stays adiabatic until $10$ pc.
The wind bubble is smaller than $5$ pc for \SI{1000}{cm^{-3}} as can be seen from the IMF-weighted thermal energy after SN. 
The decrease of the total energy between $10$ and $20$ pc for the highest density is due to energy losses as the SN becomes radiative.

Next, we compare the thermal and kinetic energy for the metallicities $Z=0.0001$ with the already discussed case of $Z=0.004$.
Relative to $Z=0.004$, the kinetic energy is similar in $Z=0.0001$ for $5$ pc and \SI{100}{cm^{-3}}. The reason is that the SN energies are similar for the metallicities and the wind bubbles all extend out to $5$ pc making the initial conditions less important. 
At the combination of $5$ pc and \SI{1000}{cm^{-3}} and $10$ pc and intermediate densities, the kinetic energy is lower at lower metallicity as SN energy is thermalised already at a smaller radius. In addition, the lack of a WR phase is reducing the kinetic energy input. 
The kinetic energy evolution with radius of the density \SI{100}{cm^{-3}} has the peculiar behaviour that it is higher at $r_\textrm{s}=20$ pc than at $10$ pc, while the thermal energy for this density shows the opposite behaviour. 
One can see how the thermal to kinetic energy ratio changes with metallicity, for example for \SI{500}{cm^{-3}} and $10$ pc the kinetic energy fraction is lower at lower metallicity.
The total energy for $Z=0.0001$ shows barely any energy loss between $10$ and $20$ pc for the density \SI{100}{cm^{-3}} as the region inbetween is completely dominated by the H\,\textsc{ii} region at the end of the stellar lifetime. 
In contrast, the total energy for the ISM density \SI{500}{cm^{-3}} decrease between $10$ pc and $20$ pc. In that density, the peak of the swept-up mass is just around $20$ pc at the end of the stellar lifetime, and the lowest mass models even exhibit densities of the order of the ambient density then. The higher densities lead to stronger energy dissipation.

As the SN energy varies with the radius for the different densities, but the injection is the same for the combinations of radius and density, the pre-SN wind is important as it sets the cavity size. 
There is less radiative cooling at lower metallicity, so the H\,\textsc{ii} region is hotter which leads to stronger feedback. Therefore there is still feedback at $20$ pc for \SI{500}{cm^{-3}}, as also SNe expand to a bigger radius in a fixed time at lower metallicity.
At $5$ pc, the total energy for nearly all cases is the same, only the thermal to kinetic ratio is different at the largest densities. It is surprising that the total energies look so similar with metallicity at $5$ pc; however,  the $5$ pc results only show how much radiative cooling one has. If one resolves $5$ pc (so cell size is around $1$ pc) one can maybe directly inject the stellar feedback (wind, photoionisation, SN) without using an intermediate step such as the 1D simulations presented here.

\subsubsection{Comparing the cumulative energy flow of all models}
\begin{figure*}
    \centering
    \includegraphics[width=1\textwidth]{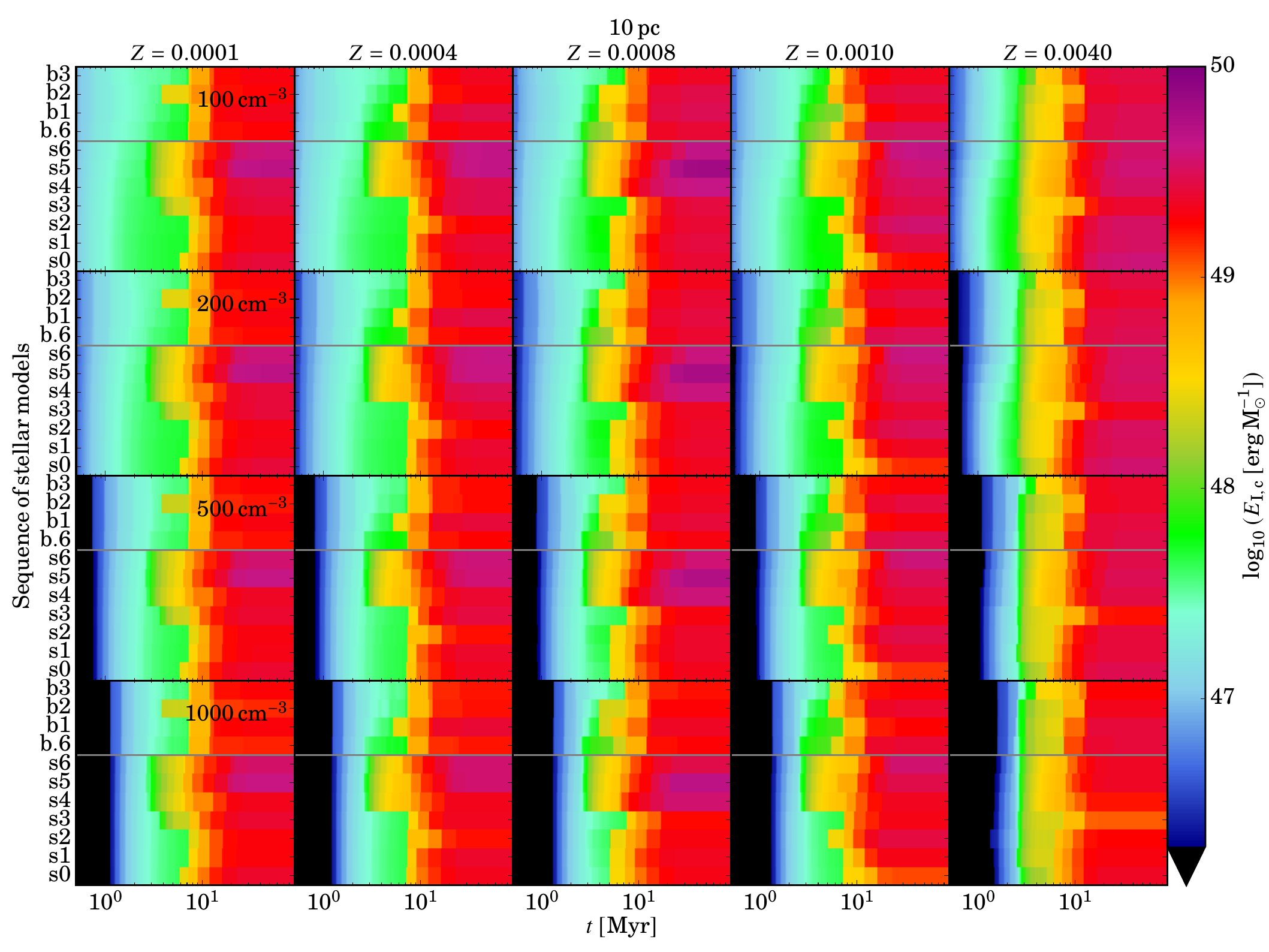}
    \caption{
    Cumulative total energy flow for all models through $r_\textrm{s}=10$ pc, weighted by an IMF (normalised to the total mass in stars above \SI{8}{M_{\sun}}). The different subpanels show different combinations of metallicity and ambient density. The grey horizontal line separates single from binary models.
    }
    \label{fig:Results_Flow_tot_10}
\end{figure*}
A comparison of the IMF-weighted total energy flows for all single and binary models can be found in Fig. \ref{fig:Results_Flow_tot_10}, with the example of the energy flux measured at $10$ pc from the star. 
The sharp energy increases visible are due to SN, and the earlier increases in same models are due to post-main-sequence phase. Additionally, a time delay until the first energy flow is visible, especially for the lower densities.
A striking result is the remarkable similarity over the whole grid of models in their total energy feedback.
This especially means that the feedback at low metallicity is comparable to the one at high metallicity due to SNe. There is no strong dependence on metallicity. Differences can be seen at early time due to the different wind strength at the different metallicities. 
However, a trend with rotation velocity is visible, with fast-rotating models that tend to have higher feedback (as they lose more mass, as we already saw beforehand). Variations in the feedback for the binary system seem to be mainly driven by the SNe.
Also in the overview, the feedback is decreasing, as expected, with the radius at which it is measured.

\begin{figure*}
    \centering
    \includegraphics[width=1\textwidth]{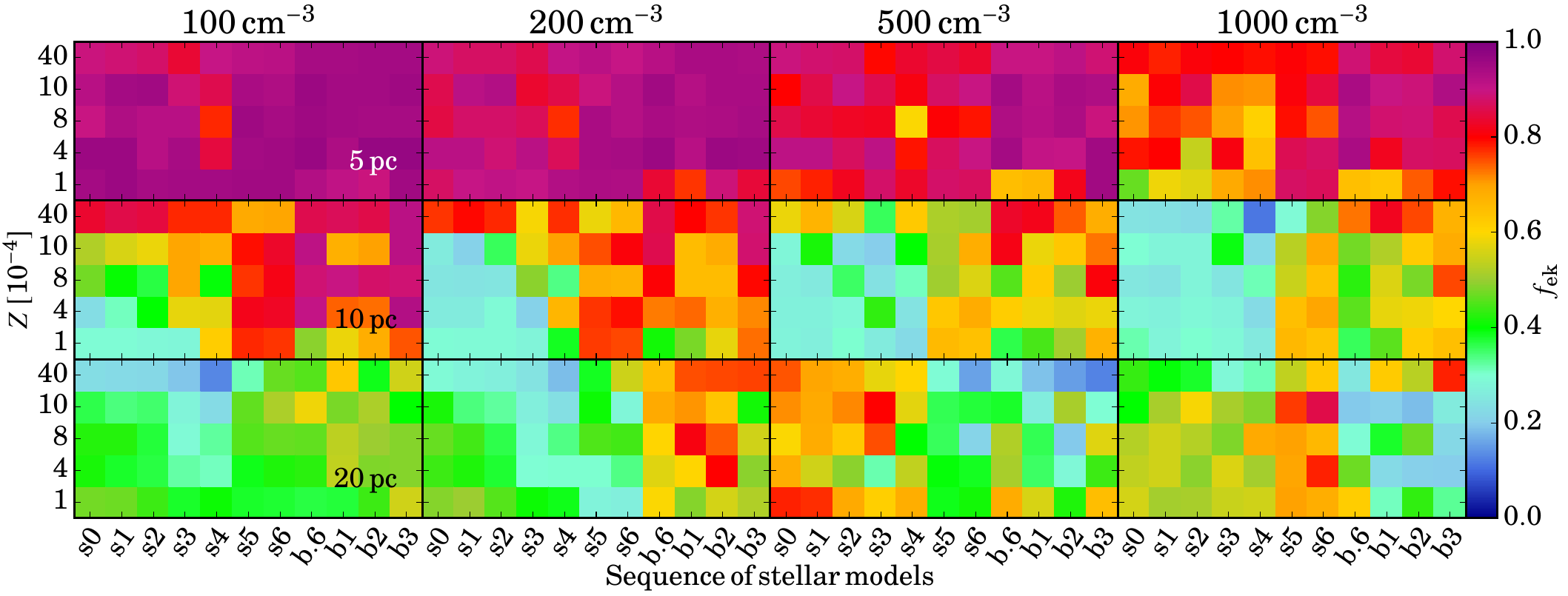}
    \caption
    {Ratio of cumulative kinetic energy flow through three $r_\textrm{s}$ values to the cumulative total energy flow at the final time.  The values are IMF-weighted. Each subpanel shows all combinations of tracks and metallicity, while the different subpanels show different combinations of ambient density and radius.  
    }
    \label{fig:Results_fek}
\end{figure*}

Regarding the fraction of cumulative kinetic energy to the cumulative total energy (see Fig.~\ref{fig:Results_fek}), 
lower radii have higher fractions as the kinetic energy gets thermalised outwards.
For $5$ and $10$ pc lower densities have higher fractions as less mass can get shock heated. Additionally, at $10$ pc a trend with the stellar rotation velocity can be seen (except for the highest metallicity). As there are more stellar models with a WR phase at higher rotation velocity, more kinetic energy is injected during the stellar lifetime. This leads to a bigger cavity, which enables more of the kinetic SN energy to be retained.
At $20$ pc, the overall energy dissipation removes trends, especially at the higher densities.
At early times, the kinetic fraction tends to be lower, while at later times SN and post-main-sequence winds lead mainly to higher kinetic energies, increasing this fraction.

\subsection{Feedback efficiency}
\begin{figure*}
    \centering
    \includegraphics[width=1\textwidth]{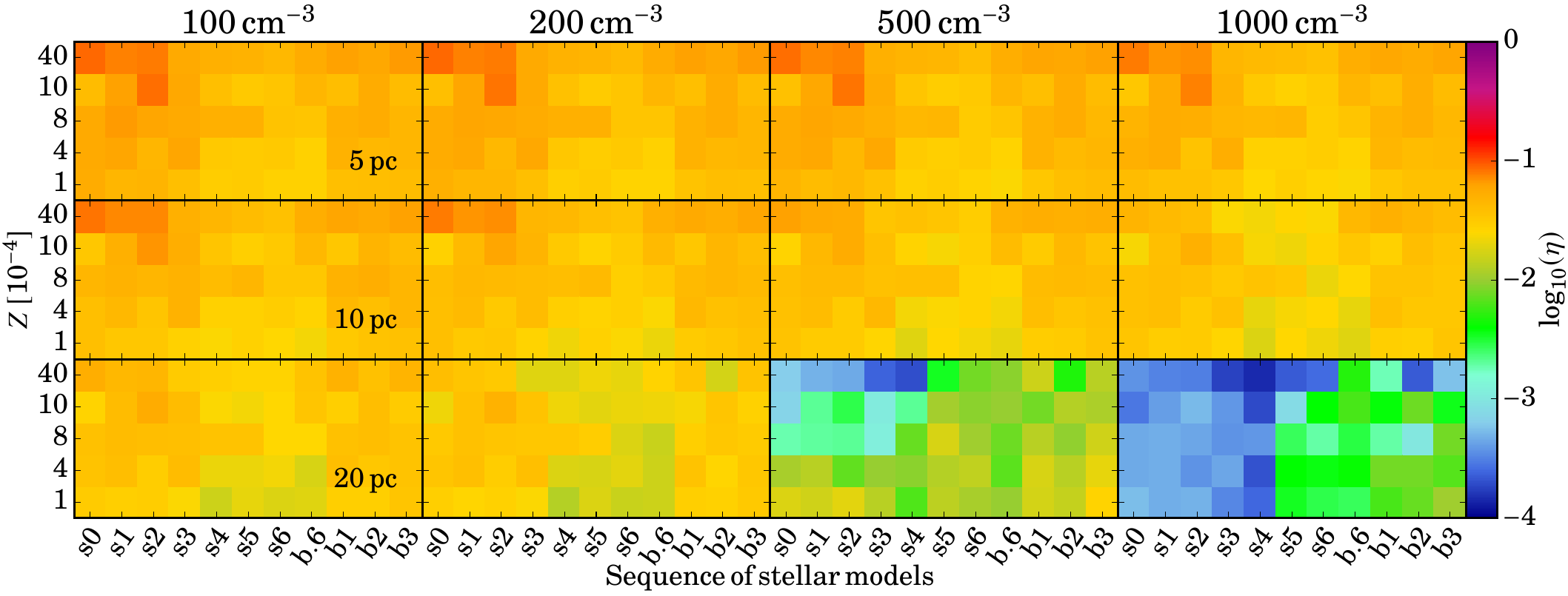}
    \caption{Ratio of cumulative total energy flow through three $r_\textrm{s}$ values to the cumulative energy input through winds, SN, and photoionisation at the final time.  The values are IMF-weighted. Each subpanel shows all combinations of tracks and metallicity, while the different subpanels show different combinations of ambient density and radius.  
    }
    \label{fig:Results_efficiency_with_ph}
\end{figure*}
The fraction of input energy (in the form of stellar winds, SN and photoionisation) flowing through $r_\textrm{s}$ between $5$ and $20$ pc, the feedback efficiency, is mostly around a few per cent, and remarkably similar for all different tracks (see Fig. \ref{fig:Results_efficiency_with_ph}). This reflects the low coupling efficiency of photoionisation. The two highest densities at $20$ pc have nearly all lower efficiencies, as radiative losses are strong. We note that the binary and fast rotating single models have an efficiency that is around one order of magnitude higher than the slower rotating single models at $20$ pc and highest density.
The metallicity $Z=0.004$ without stellar rotation has a higher efficiency as it has more SN, while the one with \SI{300}{km\,s^{-1}} has the least number of SN.
A caveat in the efficiency estimation is that the photoionisation energy can be deposited at a higher radius than $5$ pc, which is therefore not captured in the kinetic and thermal energy flow measurements.
However, it is included in the input energy estimation for all radii. 
\begin{figure*}
    \centering
    \includegraphics[width=1\textwidth]{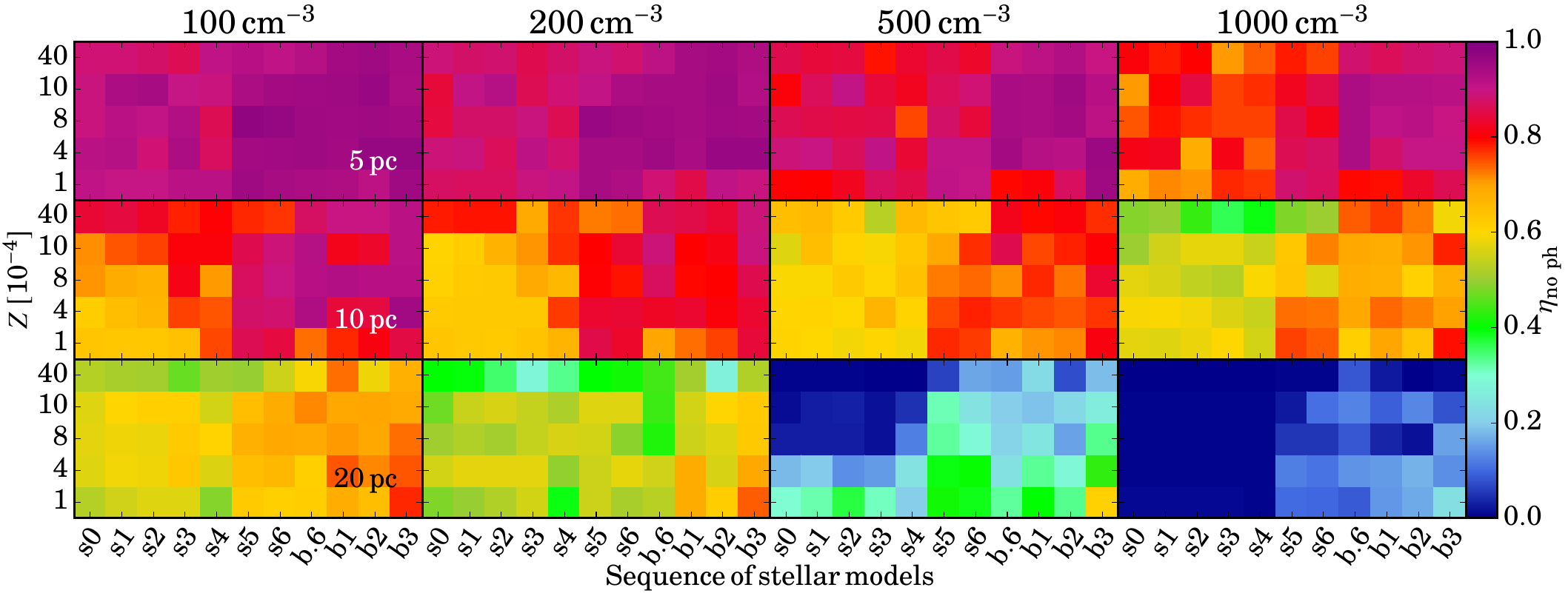}
    \caption{Same as Fig. \ref{fig:Results_efficiency_with_ph}, but the energy input only considers winds and SN, but no photoionisation. We note the different colour scale. 
    }
    \label{fig:Results_efficiency_without_ph}
\end{figure*}
We saw already that the photoionisation has such a low efficiency and the total flow at $5$ pc is following closely the input energy of winds and SN. Therefore, it is worthwhile to look at the efficiency without considering the photoionisation energy in the input (see Fig. \ref{fig:Results_efficiency_without_ph}; however,  the \textsc{Pion} simulation flows all include it). 
Considering only stellar winds and SNe is also similar to typical subresolution models of feedback in galaxy formation simulations without radiation transfer.
This efficiency therefore gives an indication of the different total energy budget available for driving feedback when injecting stellar winds and SN directly or using results from circumstellar medium simulations like these.

At \SI{100}{cm^{-3}} and $5$ pc, the feedback energy taken directly from \textsc{Mesa} are similar to those obtained with \textsc{Pion} so (nearly) all the energy is retained. This indicates that if one can resolve $5$ pc and can include photoionisation, one can directly use the results from the stellar models.
However, at the same density, but a radius of $20$ pc, only around half of the original input energy is still available. 
At higher density, the energy ratio drops further.
At \SI{1000}{cm^{-3}} and $20$ pc, especially the slower rotating single models contain orders of magnitude less energy, and binary models up to 30 per cent of the input energy. In these cases of orders of magnitude less energy, a subresolution model might just inject the remaining SN momentum.
While it would be possible, this efficiency never reaches values above one as the energies are dominated by SNe. Additionally, the photoionisation (that is neglected in this plot) is mostly not injected at the lowest radii, while at larger radii energy dissipation is already important, limiting the achievable efficiency.

\subsection{Feedback energy of stellar populations}
\begin{table}
    \centering
    \caption{Minimal and maximal cumulative energy flow and kinetic energy fractions at the final time.}
    \label{tab:Results_Population_Etot_fek}
    \begin{tabular}{ccccc}
         \hline 
            \noalign{\smallskip}
            $r_\textrm{s}$ [pc] & \multicolumn{2}{c}{$E_\textrm{tot}$ [\si{erg\,M_{\sun}^{-1}}]}  & \multicolumn{2}{c}{$f
        _\textrm{ek}$}\\
         & min & max & min & max \\ 
         \noalign{\smallskip}
         \hline 
        \noalign{\smallskip}
        $5$& \num{2.2e49} & \num{3.2e+49}  & \num{0.69} & \num{0.95} \\
         $10$& \num{1.8e+49} & \num{3.0e+49} & \num{0.44} & \num{0.86} \\
         $20$& \num{6.0e+47} & \num{2.2e+49} & \num{0.19} & \num{0.60} \\ 
         \noalign{\smallskip}
         \hline 
    \end{tabular}
    \tablefoot{Minimal and maximal cumulative energy flow through the three $r_\textrm{s}$ values for stellar populations of the considered metallicities and densities at the final time. Similarly, the minimal and maximal final kinetic energy fractions are shown. The values are IMF-weighted (normalised to the total mass in stars above \SI{8}{M_{\sun}}), and a rotation distribution for single and orbital period distribution for binary stars, and a binary fraction is assumed. As comparison, the SN-only energy (without \textsc{Pion}) for a population, if all single models in this work would undergo a CCSN with the canonical energy of \SI{e51}{erg} would be \SI{5.0e49}{erg\,M_{\sun}^{-1}}, while considering the whole grid (with binary stars) in the same way would give \SI{3.1e49}{erg\,M_{\sun}^{-1}} (see Sect. \ref{sec:Stellar_models_SN}).
    }
\end{table}
\begin{figure}
    \centering
    \includegraphics[width=1\columnwidth]{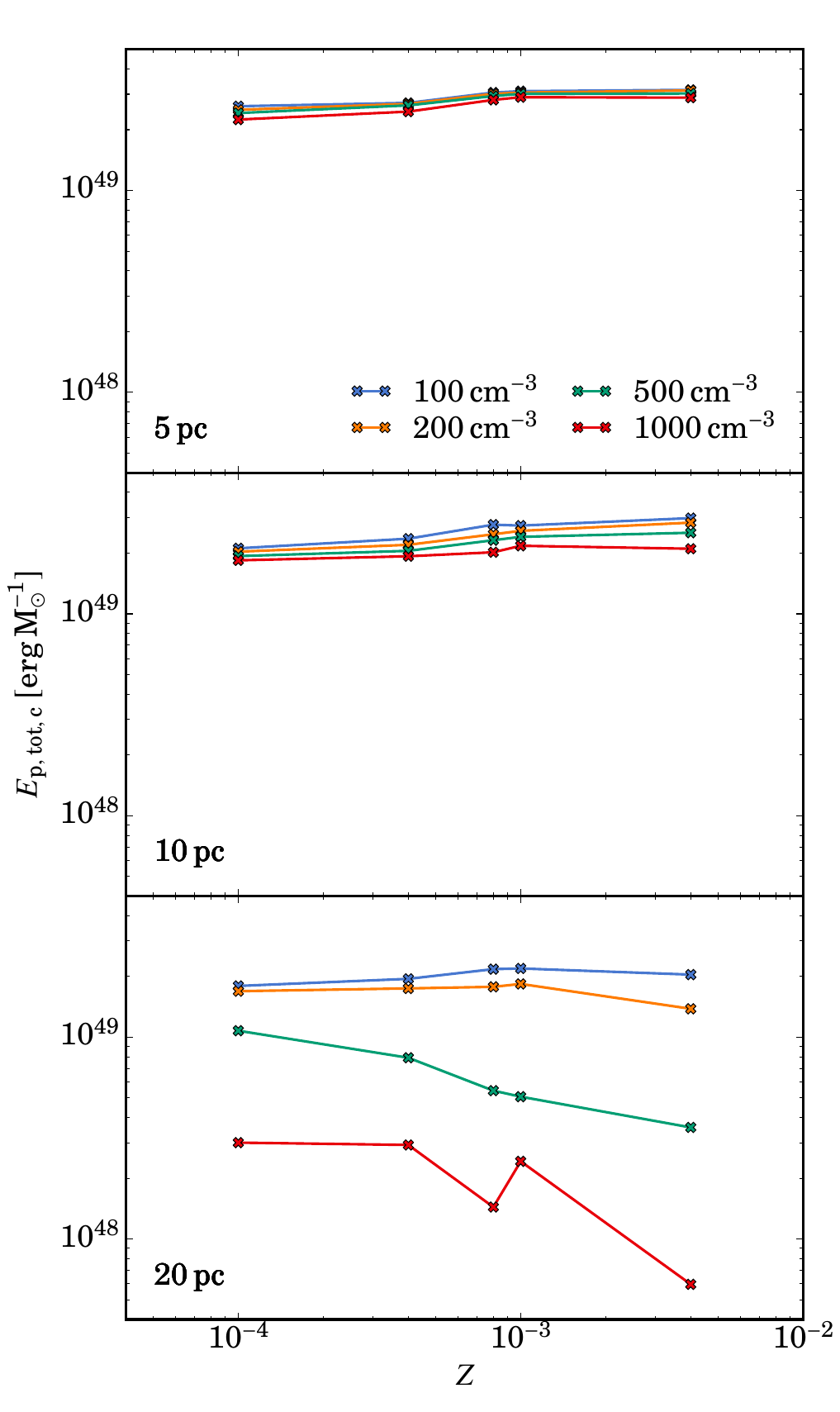}
    \caption{Cumulative total energy for stellar populations as a function of metallicity for different density and $r_\textrm{s}$. The values are IMF-weighted (normalised to the total mass in stars above \SI{8}{M_{\sun}}), and a rotation velocity distribution for single and orbital period distribution for binary stars, and a binary fraction is assumed.}
    \label{fig:Results_E_vs_Z}
\end{figure}
We can further combine the results into stellar populations. For this, we use the metallicity-dependent initial rotation velocity distribution for single stars and the orbital period distribution for binary stars of \citet{Fichtner2022}. 
The first was based on observations of the rotation velocities of young stars in the Galaxy, Large Magellanic Cloud and SMC, while the orbital period distribution was based on the results of \citet{Kobulnicky2014} and \citet{Almeida2017}.
Additionally, we assume a binary mass fraction of $70$ per cent.
The remaining parameters are then metallicity, ambient density and $r_\textrm{s}$ dependent. 

We show the ranges of the total energy and the kinetic fraction in Table \ref{tab:Results_Population_Etot_fek} for the three $r_\textrm{s}$ values.
Additionally, Fig. \ref{fig:Results_E_vs_Z} shows the cumulative population energy as a function of metallicity for the different densities and $r_\textrm{s}$ values. At $r_\textrm{s}=5$ and $10$ pc, all populations have remarkable similar total energies, and only at $20$ pc a clear density dependence can be seen.
For the highest densities at $20$ pc, the energy budget is completely dominated by the binary stars due to their order of magnitude higher retained energy compared to slow rotating single stars.
Opposite to what might be the expected behaviour, the energy is lowest for the highest metallicity at this radius and density. The reason is that cooling is stronger at higher metallicity. For the contribution of the single stars, there is the additional effect that lower metallicities contain higher fractions of faster rotating stars. 
The dip visible in Fig. \ref{fig:Results_E_vs_Z} for the highest density and radius is originating from the binary models. They have a higher energy at $Z=0.001$ than $Z=0.0008$ due to differences in the number of SNe and feedback bubble size.

The typical kinetic energy fraction decreases with radius. At $r_\textrm{s}=5$ pc the kinetic energy is for all cases dominating over the thermal energy, with the lower densities having the higher kinetic energy fractions. Additionally, a metallicity dependence due to cooling can be seen for $r_\textrm{s}=10$ pc,  where stronger cooling leads to higher kinetic energy fractions.

\subsection{Comparison to a case without pre-SN feedback}
\label{sec:Results:without_preSN_feedback}
\begin{figure*}
    \centering
    \includegraphics[width=1\textwidth]{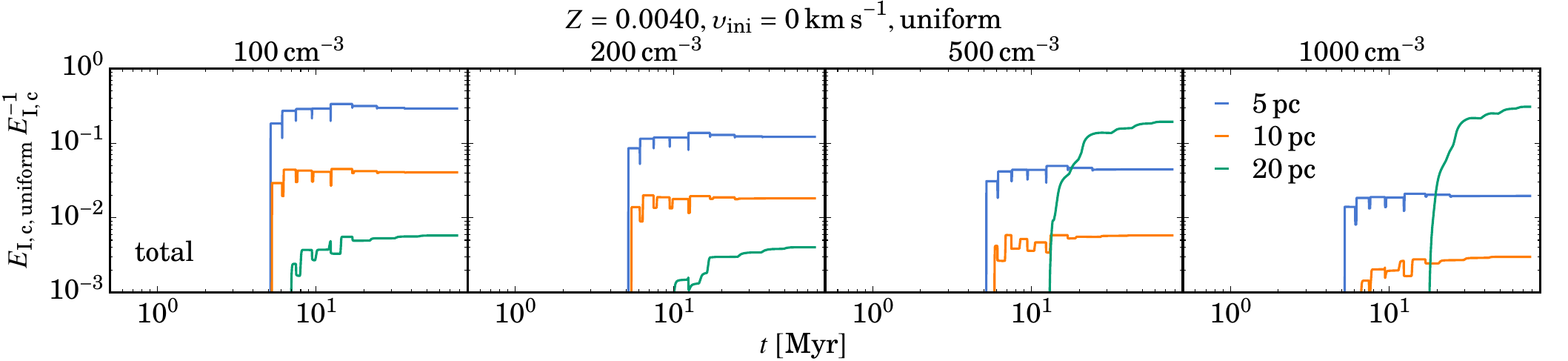}
    \caption{Ratio of cumulative total energy flow for a model without pre-SN feedback to the standard model with pre-SN feedback for single star models with $Z=0.004$ and no initial rotation velocity. The values are both IMF-weighted (normalised to the total mass in stars above \SI{8}{M_{\sun}}) and the SNe are placed at the same time as in the standard model.}
    \label{fig:Results_comparison_without_preSN_feedback}
\end{figure*}
In Sect. \ref{sec:Results:evolution_32}, we see a steep increase of the total energy at the time of the SNe, which gives rise to the question of the importance of the pre-SN feedback phase.
To investigate this, we simulate the SNe of single stars with the highest metallicity and no rotation velocity in an unperturbed medium.
Fig. \ref{fig:Results_comparison_without_preSN_feedback} shows that less energy reaches the different radii without pre-SN feedback. For most cases the differences are more than one order of magnitude,
as the low density feedback bubble helps to retain energy. 
For the lower two densities at $5$ pc, the total energy without SN is $10$ to $30$ per cent of the one in the fiducial case.
In the other two cases with a higher fraction (highest two densities at $20$ pc) also the standard model has radiated away most of the energy already. Overall, it is visible that the simulations without pre-SN feedback become radiative at a lower radius than the ones with pre-SN feedback. 
The dips in the ratio are due to the SN energy reaching $r_\textrm{s}$ earlier in the case with a cavity through pre-SN feedback than without pre-SN feedback.
The order of magnitude difference in the energy feedback when including or not pre-SN feedback clearly highlight the importance of the pre-SN feedback.


\section{Discussion}
\label{sec:Discussion}
\subsection{Comparison to previous work}
Previous studies have already examined the interaction between feedback from a single source and the ambient medium.
For instance, \citet{Fierlinger2016} investigated the efficiency of the stellar wind and SN for a \SI{60}{M_{\sun}} stellar model with solar metallicity in an ambient density of \SI{100}{cm^{-3}}. Without pre-SN feedback, they find that the retained kinetic energy, measured when the velocity of the highest density cell has decreased to the sound speed of the ambient medium, is about $0.1$ per cent of the input \SI{e51}{erg}.
Similarly, in our test without pre-SN wind at the same ambient density, albeit at a lower metallicity of $Z=0.004$, we find retained kinetic energies between $0.04$ per cent and $0.4$ per cent (increasing with increasing SN energy) at $20$ pc. 
When including stellar winds, the retained energy increases by a factor of 
around $3$ at $5$ pc, $19$ at $10$ pc and $27$ at $20$ pc for our highest mass model, which has a SN in the test case. \citet{Fierlinger2016} found an increase of the retained kinetic energy by a factor of six, which agrees with the range found in this work.
However, the comparison of our results must be done with caution, as we also include photoionisation in our models.

\citet{Freyer2003} modelled the evolution of a uniform medium with number density \SI{20}{cm^{-3}} around a \SI{60}{M_{\sun}} star at solar metallicity, affected by the feedback of wind and photoionisation. 
Due to a different treatment of the photoionisation energy, we recalculate their efficiency, the ratio of energy in their domain to the cumulative input, according to the method presented for the photoionisation energy in Sect.~\ref{sec:Results:overview_energy_flow} by using their average stellar temperature and luminosity. With this method, the efficiency in \citet{Freyer2003} is around \num{1.4e-2} at the end of the stellar lifetime.
The efficiency of the cumulative energy flow of our most similar model, $\log_{10}(M/\mathrm{M}_\odot)=1.8$ without stellar rotation, is around \num{9.8e-3} at a distance of \SI{5}{pc} at the end of the stellar lifetime. This is lower but within a factor of two of \citet{Freyer2003}. However, our simulation has a lower metallicity of $Z=0.004$ and a higher ambient density of \SI{100}{cm^{-3}}. Fig.~\ref{fig:Results_efficiency_with_ph} displays a weak trend of higher efficiency for higher metallicities and lower densities, albeit for whole stellar populations with SNe. Considering this, we find no disagreement between the efficiency of our model and the one of \citet{Freyer2003}.

\subsection{Uncertainties and caveats}
Despite the computational effort, uncertainties remain in the models used in this work. 
Among these are the stellar modelling \citep[for a discussion see][]{Fichtner2022}, where for instance the rate of mass-loss is still under debate \citep[e.g.][]{Bjorklund2023, Sabhahit2022}, the coupling of the stellar models into \textsc{Pion}, the SN rate and energy, the composition of stellar populations and the 1D simulations with \textsc{Pion}. 
Spherically symmetric simulations cannot capture turbulence or its dissipative effects through mixing hot and cool gas phases, which is especially effective at the wind-ISM interface \citep{MacGvaMoh15, Lancaster2021}.  Approximate methods have been applied to the expansion of superbubbles driven by SN feedback \citep{ElBadry2019}, but these should always be calibrated to 3D simulations and observations, and in addition the metallicity/magnetic-field dependence of such schemes is unclear. 

\citet{Geen2015} found that 3D simulations with uniform initial conditions result in a circumstellar medium with almost spherical symmetry and they do not expect large differences compared with 1D simulations with similar initial conditions, but a comparison of feedback efficiency for 3D vs.\ 1D was not carried out.
We cannot compare our results meaningfully with \citet{Geen2015} because of the different stellar mass and metallicity.
The \citet{Geen2015} simulations produced spherically symmetric structures because (i) the initial conditions were uniform with uniform external pressure and (ii) the \SI{15}{M_{\sun}} star in their calculations only has main sequence and red supergiant phases, during which strong dynamical instabilities do not develop for expansion into a uniform ISM.
\citet{Arthur2011} showed that substruture is diminished for H~\textsc{ii} regions expanding around B stars compared to more massive O stars, because they emit a softer radiation field, producing a weaker temperature jump between ionised and neutral gas.
While this and other diffusive processes may destroy substructure in circumstellar nebulae and favour a more spherically symmetric evolution \citep{McKee1984, Arthur2011}, there are also physical processes and environmental factors that drive the solution away from spherical symmetry.
Simulations of a more massive star than used in \citet{Geen2015} with wind-wind interactions from a WR phase would have resulted in the development of strong dynamical instabilities breaking spherical symmetry, as in \citet{Freyer2006}.
Further deviations from spherical symmetry can be caused, for example, by a structured ISM, the movement of the star providing feedback \citep[e.g.][]{MacLow1991,MacGvaMoh15}, turbulent clouds \citep[e.g.][]{Medina2014}, self-gravitating clouds \citep[e.g.][]{GeenHennebelle2015}, stellar rotation \citep{Chita2008} or the rotational and gravitational field of the cloud \citep[e.g.][]{Dalgleish2018}.

In particular, the multiphase or stratified, non-uniform density structure of the ISM can lead to varied expansion rates and efficiencies of feedback bubbles, which could allow the feedback bubble to expand further during the stellar lifetime,
resulting in larger radii than shown in Fig.~\ref{fig:Results_Spitzer_comparison}. These effects of ISM structures and dynamical instabilities are expected to be stronger for higher mass stars with more intense ionising radiation fields \citep{Arthur2011}.
This pre-SN bubble size is important for the radius at which the SN feedback efficiency drops.
Fractal clouds with strong density stratification have been shown to increase the effectiveness of SN feedback \citep[e.g.][]{WalchNaab2015, Haid2018}, and analytic models and simulations show that such structures also influence pre-SN circumstellar nebulae \citep{McKee1984, Rogers2013, Haid2019}.
The growth of the H\,\textsc{ii} region can be further influenced, for instance through champagne flows, leading to changes in the detailed evolution (e.g. of Fig. \ref{fig:Results_Spitzer_comparison}).
H~\textsc{ii} region expansion into power-law density profiles can differ from expansion into a uniform medium through the development of ionisation-front instabilities \citep{GarciaSeguraFranco1996}.
Inclusion of stellar winds further complicates the hydrodynamics of circumstellar material in 3D simulations \citep[e.g.][]{Geen2023}.

It is not possible to capture these inherently multidimensional effects with our 1D simulations because of the imposed symmetry.
Future work making quantitative comparison between 1D and 3D solutions for more realistic environments would be valuable to assess the accuracy of our approach to subresolution feedback modelling.
In this aspect, the presented results may be seen as lower limits of the possible feedback for $r_\textrm{s}$ values higher than the pre-SN bubble radius.
We note, however, that our 1D approach represents an effort to improve on current subgrid implementations of stellar feedback in simulations of galaxy evolution.
These implementations are effectively `0D' in the sense that stellar winds, radiation and SNe are used directly from stellar evolution calculations.

Furthermore, each of our 1D simulations includes only one source (a single star or a binary system), whereas star formation is often clustered. Clustering could alter the large-scale impact of pre-SN and SN feedback. Multiple stars in close proximity could drive a wind bubble and H\,\textsc{ii} together, creating a larger cavity \citep{MacLow1988, rahner2017}. This could allow more SNe to provide efficient feedback and the feedback energy to be retained until larger radii.
Superbubbles created by the correlated feedback of stellar populations can escape out of the galactic disc \citep{MacLow1988}. This can launch galactic fountains and outflows, through which energy, mass and metals can be transported into the circumgalactic medium and beyond. We plan to use the results of our 1D calculations in a subgrid model for stellar feedback to investigate galactic winds in large-scale simulations of galaxy formation.

We do not consider SN Type Ia in this work. They originate from lower mass stars and explode with a large time delay after the stellar birth. Due to this, the explosion does not occur close to the birth place. Therefore, their evolution is not strongly influenced by the early feedback of massive stars and can be modelled separately.

\subsection{Consequences of our results for stellar feedback}
In nearly all cases for the considered parameters, one star or binary system is able to create a cavity of at least \SI{10}{pc} size. 
Similarly, \citet{Chevance2020} found that a populations of stars can disperse giant molecular clouds already in the first few Myr after the first massive stars form.
The importance of the post-main-sequence phase is in agreement with the findings in the superbubble N\,206 in the SMC. In this low-density region, one WO binary is completely dominating the pre-SN feedback budget over the $320$ OB stars \citep{Ramachandran2019}.

Rotating and binary systems are thought to be important for galaxy formation due to their higher feedback. While the feedback energy from stellar winds of a population of rotating and binary stars of our grid was found to be around one order of magnitude higher than from single non-rotating stars at the lowest metallicity of $Z=0.0001$, the global properties of a galaxy simulated with a feedback subgrid model based on these different populations were remarkable similar \citep{Fichtner2022}. Consistent with that, we find, in our 1D simulations of the circumstellar medium, 
similar energies (within a factor of two) for a population of single non-rotating stars, rotating single stars, binary stars and a population combining single and binary stars when evaluated for the lowest density at $r_\textrm{s}=5$ pc. This suggests that the importance of rotating and binary stars might be overestimated. 
It is enough if the pre-SN feedback is able to create a cavity, and then the SN energy input is the main contributor to feedback energy reaching larger scales.
Contrary to expectations, main-sequence winds of massive stars do not dominate the pre-SN feedback; the main contributors are post-main-sequence winds and photoionisation.

Populations of binaries, but also some populations of single stars with rotation velocities of \num{500} or \SI{600}{\kilo\meter\per\second}, retain around one order of magnitude more feedback energy than single non-rotating models at $r_\textrm{s}=20$ pc for the highest density (see e.g. Fig. \ref{fig:Results_efficiency_with_ph}). In these populations, the feedback of some stellar models is able to evacuate the star's surrounding to at least \SI{20}{pc} before the (secondary) SN occurs. For the population of binaries, the models with two SNe with a primary mass above $10^{1.4}$ M$_\odot$ are of particular importance to create such large cavities. Therefore, the different strength of pre-(secondary)-SN feedback is mainly important through the size of the cavity. 
However, this is mainly relevant when measuring the energy at a $r_\textrm{s}$ similar to the cavity size. Far inside the cavity the similar SN energies lead to similar cumulative energies flows, while far outside the energy dissipation by cooling dominates.

The feedback energy return is nearly independent of metallicity. At large radii even an opposite trend than expected can be seen:  the energy feedback is higher at lower metallicity due to the weaker metal cooling. This trend might hold also true for lower metallicity. However, the question arises if in a metal-free, population III setting a similar behaviour is possible. To approximate such a situation, we rerun one simulation of a non-rotating \SI{32}{M_{\sun}} stellar model from the lowest metallicity without any cooling from metals and no stellar wind \citep{Krticka2006,Krticka2009}. 
We find that the energy at $r_\textrm{s}=5$~pc is lower with \SI{1.2e51}{erg} compared to the fiducial case with \SI{1.7e51}{erg}, likely due to the missing stellar winds. This is still higher than the \SI{7.7e50}{erg} of the SN-only test case (which has, however, a higher metallicity and therefore another, but higher, SN energy). In contrast, at $r_\textrm{s}=20$~pc, the feedback energy of the population III-test case is higher than the fiducial case, with \SI{7.0e50}{erg} to \SI{5.2e50}{erg}, while the SN-only test case has radiated away nearly all the energy.
The feedback from photoionisation and SN in a metal-free ISM is able to create a larger feedback bubble than in the fiducial case, leading to the higher energy at $20$ pc.
The lower energy injection due to the lack of stellar winds is (over-)compensated by a lower energy dissipation due to less cooling.
Therefore, even for population III stars the pre-SN feedback might be able to create a feedback bubble that is able to support a SN energy return to scales of tens of pc. This might reduce SF in the surroundings of the first stars \citep[][]{Mackey2003} and expel the gas from the low-mass host minihalo \citep{Klessen2023}.

Our results show that the details of radiative and mechanical pre-SN feedback are not crucial to determining the overall feedback energy returned to the ISM, but the primary dependence is on which stars explode and with how much energy. 
We use the compactness parameter to determine explodability and explosion energy following \citet{O'Connor2011} and \citet{Schneider2021}, but recently \citet{Wang2022} found that their 2D SN explosion simulations disagree quite strongly with this prescription in some aspects.
In particular most of their stars explode successfully regardless of the compactness parameter and, for the more massive stars, a compact core is actually required to successfully initiate an explosion.
\citet{Burrows2024} have started a similar analysis of a suite of 3D CCSN simulations, finding similar results.
Given the rapid development of this field and current lack of consensus in the literature, this must be seen as one of the largest uncertainties in determining feedback from stellar populations.

As well as these recent developments in SN modelling, the topic of winds from massive stars and the upper mass limit has received renewed attention \citep{vink2021r}.
In particular, as very massive stars approach the Eddington limit, their mass-loss rates should increase more strongly with mass \citep{Sabhahit2022} than standard scaling laws for O and B stars.
\citet{Higgins2023} showed that if the upper mass limit is on the order of 500\,M$_\odot$ then the mass and metal yields from winds of very massive stars ($>100$\,M$_\odot$) may contribute significantly to the total, even when integrated over an IMF (albeit at high metallicity).
We have not included stars with mass $M>160$\,M$_\odot$ in this work, nor stellar models representative of binary mergers, but it will be interesting to see what effect such stars would have on the cumulative energy feedback rates presented here.

\subsection{Implications for a subgrid feedback model}
\begin{figure}
    \centering
    \includegraphics[width=1\columnwidth]{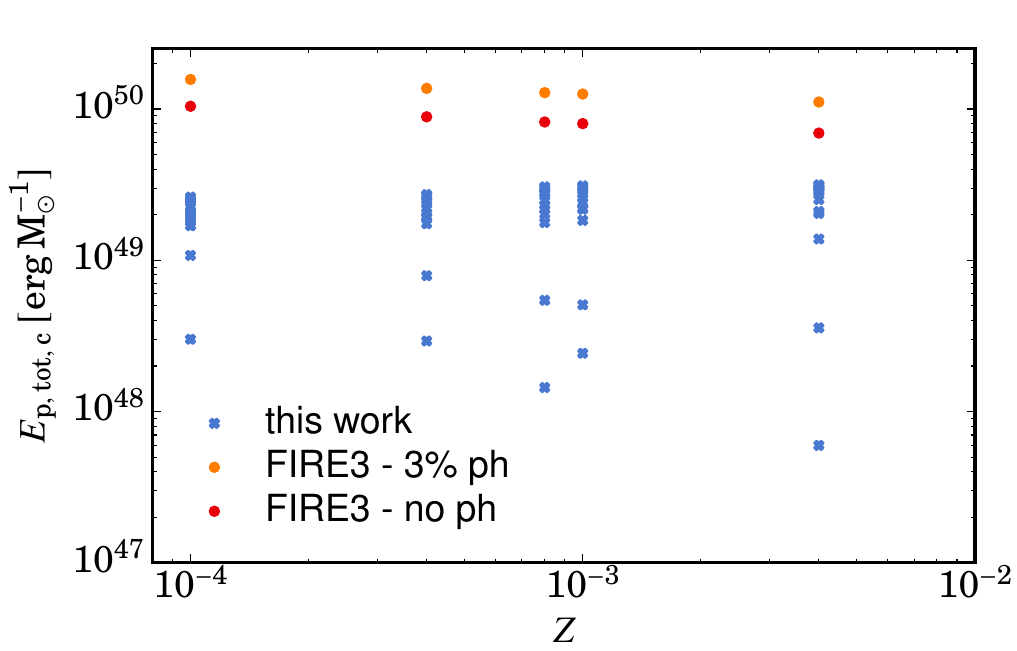}
    \caption{Feedback energy from a stellar population as a function of metallicity for different density and radii is shown with blue cross symbols. 
    The red dots indicate the energy feedback from CCSNe and stellar winds in \textsc{FIRE-3} \citep{Hopkins2023}, while the orange dots include additionally photoionisation feedback that is assumed to have a coupling efficiency of $3$ per cent.
    All values are IMF-weighted (normalised to the total mass in stars above \SI{8}{M_{\sun}}).
    }
    \label{fig:Discussion_comparison_fire3}
\end{figure}
The use of 1D simulations of the interaction with the surrounding medium leads to changes compared to taking values directly from stellar models. The first difference is a lower feedback energy due to energy dissipation on scales that are not resolved. For example, in cosmological simulations this  might lead to an overestimate of the effect of stellar feedback on a galactic scale.
In addition, there is a time delay between the ejection of the stellar feedback and its arrival at distances further away from the star. Furthermore, the ambient medium alters the kinetic-to-total energy ratio, which decreases with increasing distance from the source.

As discussed earlier, we find nearly no metallicity dependence of the feedback energy. This has implications for high-redshift or low-mass galaxies which have a low metallicity, as it increases the impact of feedback on these systems. 
The inclusion of photoionisation and stellar winds, especially post-main-sequence winds, is important for feedback schemes, as they change the timing and efficiency of energy transfer to the ISM.
The energy released by SNe varies among different populations, such as populations with lower metallicity. This is particularly important as the SN energy dominates the energy budget in the circumstellar medium.

We compare our results to the widely used Feedback In Realistic Environments (FIRE) project \citep{Hopkins2023}.
To derive their feedback energy, we integrate their equations (1), (2), (4), and (5) in their Sect. 3.3 until \SI{1}{Gyr}. We then renormalise the IMF stated in their Sect. 3.3 to include only massive stars above \SI{8}{M_{\sun}}, to match the population definition used in this work.
Our feedback energy is $30$ to $100$ times lower for the highest density and largest distances, due to energy dissipation in the ambient medium (see Fig. \ref{fig:Discussion_comparison_fire3}).
Even at low density and distance, where we do not see strong energy dissipation, our energy feedback is a few times lower. This could originate from the differences in the treatment of CCSNe. In \citet{Hopkins2023}, all massive stars (up to \SI{120}{M_{\sun}}) are assumed to undergo a CCSN, while in our work only some models undergo a CCSN depending on the compactness parameter of the stellar model.
The stars around $8-10$\,M$_\odot$, which are the most numerous, have the lowest-energy explosions when using our SN prescription, 
whereas \citet{Hopkins2023} assume that the explosion energy is independent of mass.

Subgrid models of SN feedback commonly inject energy only in one component, for example thermal energy, or keep a fixed ratio between different components.
However, we unveil a much more complicated picture. 
The fraction of energy in the kinetic component is a complex function of time, 
evolutionary phase, metallicity, density and distance, which should be accounted for to create more realistic subgrid models of stellar feedback.

The density and resolution in galaxy formation simulations are connected to each other, in the sense that higher resolutions enable higher densities.
For instance, in large-scale simulations like IllustrisTNG-100 gravitational softening lengths of the order of \SI{100}{pc} with star formation thresholds of \SI{0.1}{cm^{-3}} are reached \citep{Pillepich2018,Pillepich2018b}, while recently zoom-in simulations of \citet{Gutcke2022} use a gravitational softening of $0.5$ pc and $4$ pc for gas and stars, respectively and a star formation threshold of \SI{1000}{cm^{-3}}. 
For the available feedback budget, we find that a higher density and a larger measurement radius, $r_\textrm{s}$, have a similar effect: they both lead to more energy dissipation. Therefore, the impact of larger distances (i.e. lower resolution) in galaxy simulations could be reduced by the corresponding lower achieved densities and might help to maintain a higher feedback energy budget despite the lower resolution. 

Implementation of our feedback prescription into detailed galaxy-formation simulations are required to assess the impact of the gas-altered feedback on a whole galaxy,  which may depend on the resolution adopted in the galaxy formation simulation. For resolutions of pc-scale the effect will likely be smaller than for lower resolutions, because the energy dissipation and time delay are small.
We plan to implement a subresolution model with the feedback from our 1D simulations to test against standard feedback implementations.

\section{Summary}
\label{sec:Conclusions}
Simulations of galaxy formation rely on the inclusion of stellar feedback through subgrid models, where the feedback energy is often calculated directly from the stellar ejection.
However, the circumstellar medium can alter the feedback ---with respect to that injected by stars--- on scales below the resolution of the simulations.
In this work, we investigate the resolution gap between stellar and galaxy simulations by quantifying 
the energy that reaches pc to tens of pc scales. 

We compute a large grid of simulations of the  circumstellar medium with the code \textsc{Pion}, including the feedback from stellar winds, photoionisation, and SNe of massive stars.
We adopt an extended version of the detailed grid of stellar evolutionary models presented in \citet{Fichtner2022}, which are computed with \textsc{Mesa}.
The grid includes different initial rotation velocities for single stars and orbital periods for binary systems. We use four different ambient densities and five different metallicities.
For the SNe, we perform computations to identify which models undergo a CCSN according to the compactness parameter, a commonly used criterion for explodability. 
Additionally, we assume a relation between SN energy and the compactness parameter. 

In the grid of \textsc{Pion} simulations, we evaluate the thermal and kinetic energy flowing through spheres of three exemplary radii: $5$, $10$, and $20$ pc. 
Assuming a standard Salpeter IMF for stars of 8 to 160\,M$_\odot$, our findings can be summarised as follows:
\begin{enumerate}
\item The use of the criterion for explodability, as opposed to the 
standard approach of injecting \SI{e51}{erg} per massive star, leads to a metallicity-dependent SN energy return. 
The kinetic energy input of stellar winds decreases sharply with metallicity, especially for non- or slowly rotating single stars. Despite the metallicity dependence, nearly all the stars considered are capable of producing a cavity of at least \SI{10}{pc} in radius, even at low metallicity.
\item The location of the shell of ISM material swept up by photoionisation feedback at the end of the stellar lifetime is independent of the stellar initial mass (except for the lowest mass models) in a population where all other parameters are the same.
\item The energy budget at all radii is dominated by the SN energy.
Despite this, pre-SN feedback strongly increases the efficiency of SN feedback (mostly by more than a factor of $10$) when comparing the fiducial simulations to test runs without pre-SN feedback.  The amount of retained SN energy is strongly influenced by the size of the pre-SN feedback cavity. These findings highlight the importance of pre-SN feedback and the necessity of their inclusion.
Although we treat each star in isolation in the 1D simulations, the effect of stellar clustering will increase the efficiency of energy transfer to the ISM because of the larger cavities formed.
\item The time when the feedback energy front crosses $20$ pc is significantly later than the time it is ejected from the star.
The length of this delay is significant with respect to the cloud-formation timescale. The direct injection of feedback at the time of the stellar ejection on a scale of tens of pc could therefore halt SF too early.
\item Stellar masses of around \SI{30}{M_{\sun}} make the greatest contribution to the total energy, as they are massive enough for strong winds, but still explode as CCSN, unlike more massive stars. Additionally, such stars are not as rare as the most massive stars.
\item Overall, the main sequence stellar winds are of limited importance for the energy budget. Before SN, the post-main sequence winds dominate the kinetic energy, and photoionisation dominates the thermal energy.
\item The strongest dependence of the energy feedback return is on the radius at which it is measured, with higher radii retaining lower energies. Additionally, a decrease in energy with increasing density is found, as expected. The feedback available for a subgrid model also  therefore depends on the properties of the galaxy simulation (resolution, maximum resolved gas density), suggesting that feedback models should be developed that take a larger number of dependencies of the energy into account.
\item The total feedback energies are similar for single stars with different rotation velocities and binary systems with different orbital periods, although fast-rotating single stars have the highest feedback energies overall. Especially noticeable is the near independence of the feedback energy on metallicity, even though stellar winds are strongly metallicity dependent. This is explained by the similar SN energies across the grid and by the fact that almost all stellar models do produce some stellar wind bubble and H~\textsc{ii} region cavities that allow SNe to propagate to larger radii without losing significant energy. Binary systems with two SNe form bigger cavities, leading to higher retained energies even at $20$ pc.
\item The efficiency of the feedback (the retained energy as a fraction of the input) is around a few per cent, except for the higher densities at $20$~pc, where nearly all the energy is radiated away. The overall low efficiency reflects the low coupling efficiency of the photoionisation energy input. When taking the ratio without including the photoionsation energy in the input energy, this energy is nearly completely retained at $5$ pc and around \SI{100}{cm^{-3}} (and probably below). 
\item The fraction of energy in the kinetic component is not a constant ---as assumed in standard subgrid models--- but depends on the time, ambient density, metallicity, radius, and the considered stellar track.
When assuming a binary mass fraction of $70$ per cent (with a rotation velocity and an orbital period distribution), the fraction of kinetic to total energy ranges between $0.66$ and $0.95$ close to the star, while further away the thermal energy becomes more important (kinetic fractions of $0.22$ to $0.62$) due to shock heating.
\end{enumerate}

Despite its lower energy, pre-SN feedback is essential even at low metallicities (e.g. like those found in high-redshift galaxies) as it preprocesses the star's surroundings before the succeeding SN. 
We do not find a strong decrease in the feedback energy with decreasing metallicity, 
because all calculated models produce sufficient pre-SN feedback that SNe explode into a low-density cavity and therefore remain in the adiabatic phase for longer.
Our work shows that it is not the amount of pre-SN feedback energy that is of major importance, but rather the feedback mechanism implemented, and notably the existence of pre-SN feedback, binary systems with two SNe, and the timing of the feedback. 
The ambient medium influences, for example, the energy budget, feedback delay, and the kinetic energy fraction already at scales that are subresolution in galaxy formation simulations. These effects should be accounted for in order to create more realistic numerical models of stellar feedback.

The scripts to generate the \textsc{Pion} simulations as well as the flow data of the shown simulations at the three discussed radii are publicly available.

\begin{acknowledgements}
YAF, ERD and CP gratefully acknowledge the Collaborative Research Center 1601 (SFB 1601 sub-projects C5 and C6) funded by the Deutsche Forschungsgemeinschaft (DFG, German Research Foundation) – 500700252. 
This work was also carried out within the Collaborative Research Centre 956, sub-project C04, funded by the Deutsche Forschungsgemeinschaft (DFG) – project ID 184018867. 
YAF is part of the International Max Planck research school in Astronomy and Astrophysics and guest at the Max-Planck-Institute for Radio Astronomy.
JM acknowledges support from a Royal Society-Science Foundation Ireland University Research Fellowship (20/RS-URF-R/3712).
JM is grateful to Prof.~N.~Langer for the use of a guest office at the AIfA during the time when this research was carried out.
CP is grateful to SISSA, the University of Trieste, and IFPU, where part of this work was carried out, for hospitality and support.
In addition to the software cited in the main text, this research made use of: the \textsc{Silo} library for data I/O\footnote{\href{https://software.llnl.gov/Silo/}{https://software.llnl.gov/Silo/}}, the \textsc{Sundials} ODE solver library \citep{hindmarsh2005sundials, gardner2022sundials}, \textsc{pypion} \citep{green2021pypion}, \textsc{NumPy} \citep{Numpy_2020}, \textsc{matplotlib} \citep{Hunter_2007}, \textsc{SciPy} \citep{Virtanen2020}, \textsc{Astropy} \citep{AstropyCollaboration2022} and \textsc{seaborn} \citep{Waskom2021}. 
\end{acknowledgements}

\section*{Data Availability}
The data underlying this article are available in Zenodo at doi:\href{https://zenodo.org/doi/10.5281/zenodo.13285148}{10.5281/zenodo.13285148}.


%
\bibliographystyle{aa} 
\bibliography{Ref} 
%



\begin{appendix} 
\section{Evolution of a binary system}
\label{sec:Appendix_A}
\begin{figure}
    \centering
    \includegraphics[width=1\columnwidth]{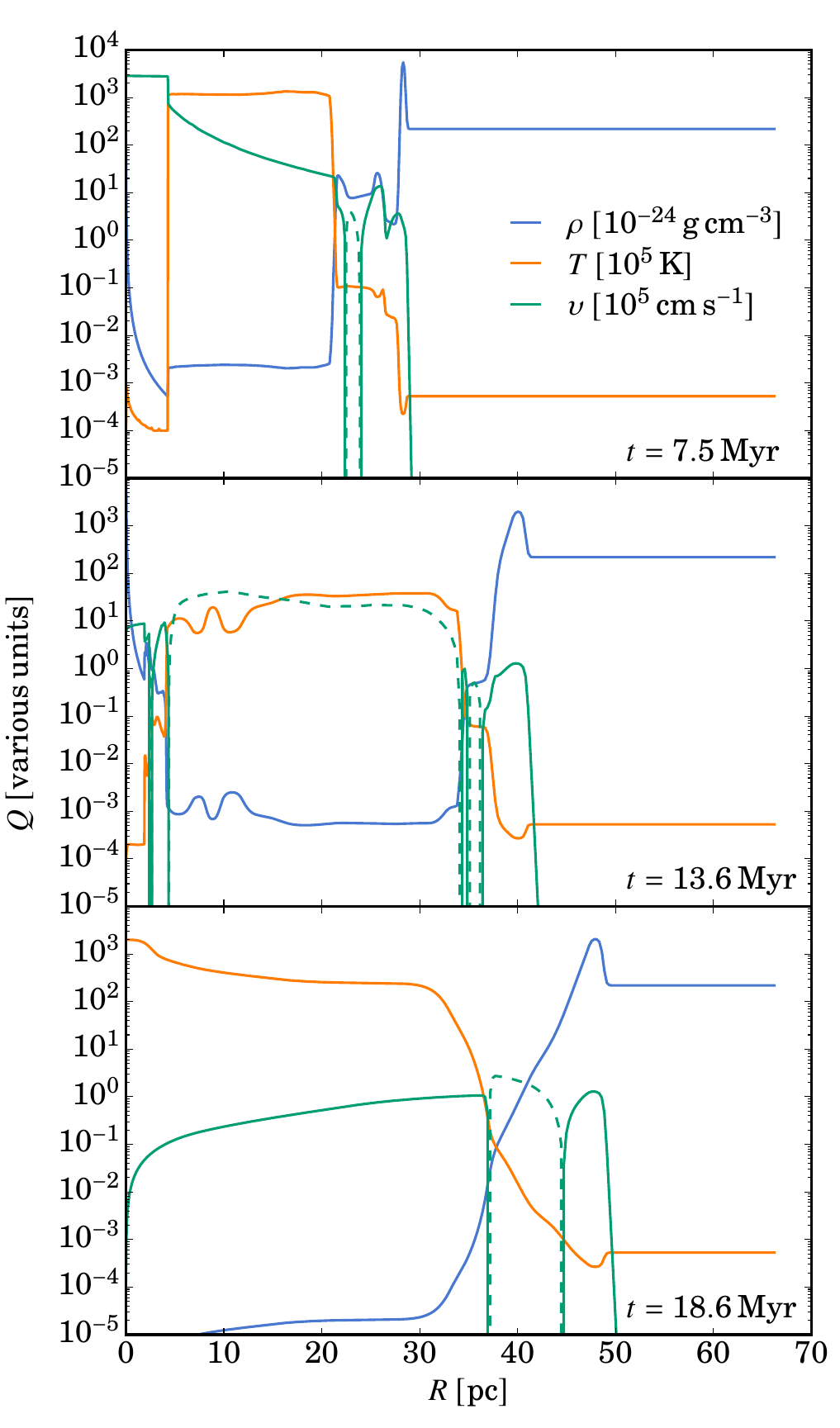}
    \caption{Density, temperature and velocity profile of a \textsc{Pion} simulation with a binary system of $Z=0.004$, $\log_{10}(P\mathrm{\,[days]})=1.0$ and $\log_{10}(M_\textrm{p} \mathrm{\,[M}_\odot\mathrm{]})=1.4$, in a uniform medium of $n=\SI{100}{cm^{-3}}$ at the end of the primary lifetime (top), at the end of the secondary lifetime (middle) and \SI{5}{Myr} after the secondary SN (bottom). Dashed lines indicate negative velocities.}
    \label{fig:Appendix_Profile_binary}
\end{figure}
We show the profile around a binary system that has two SNe (from the primary and secondary star) during the evolution in Fig. \ref{fig:Appendix_Profile_binary}, in comparison to the single star case in Fig. \ref{fig:Results_Profile_32Msun}. The mass-loss of the secondary star prevents the primary SN bubble from reaching as low densities as in the single case or after the secondary SNe (middle panel). Furthermore, the central region does not have the lowest density but rather the highest densities inside the bubble. As for the single star case, the secondary SN removes all structures inside the bubble (lower panel). 
The feedback bubble extends out to nearly $50$ pc \SI{5}{Myr} after the secondary SN, while in the case of the corresponding single star only $40$ pc are affected \SI{5}{Myr} after the SN. This implies that the volume of the cavity by the binary SNe is twice as big as in the single case.

\end{appendix}

\end{document}